\documentclass[longauth]{aa}

\usepackage{graphicx}
\usepackage{natbib}
\usepackage{scalerel}
\usepackage{comment}
\usepackage{multirow}	

\usepackage[table]{xcolor}

\usepackage{txfonts}
\usepackage[pdfencoding=auto,psdextra]{hyperref}
\hypersetup{
    colorlinks=true,
    linkcolor=blue,
    filecolor=magenta,      
    urlcolor=blue,
    citecolor=blue
}
\makeatletter
\renewcommand*\aa@pageof{, page \thepage{} of \pageref*{LastPage}}
\makeatother

\usepackage[utf8]{inputenc}

\usepackage[switch, modulo]{lineno}
              
\renewcommand{\linenumbers}[0]{}

\usepackage{euclid}

\newcommand{\DEMNUni}{\texttt{DEMNUni}\xspace}

\newcommand{\DUSTGRAIN}{\texttt{DUSTGRAIN}\xspace}
\newcommand{\DUSTGRAINPATHFINDER}{\texttt{DUSTGRAIN-PF}\xspace}
\newcommand{\rockstar}{\texttt{Rockstar}\xspace}
\newcommand{\lcdm}{$\Lambda$CDM\xspace}

\newcommand{\ol}{\Omega_\Lambda}

\begin{document}
%
% Put the title of your paper here:
%
\title{\Euclid preparation}
\subtitle{The shape of halo profiles in $\Lambda$CDM and non-standard cosmologies}    

\newcommand{\orcid}[1]{} %% if already defined in aa.cls: comment, or use renewcommand			   
\author{Euclid Collaboration: L.~Pizzuti\orcid{0000-0001-5654-7580}\thanks{\email{lorenzo.pizzuti@unimib.it}}\inst{\ref{aff1}}
\and G.~Y.~Ferron\inst{\ref{aff1}}
\and A.~Ragagnin\orcid{0000-0002-8106-2742}\inst{\ref{aff2}}
\and A.~M.~C.~Le~Brun\orcid{0000-0002-0936-4594}\inst{\ref{aff3}}
\and P.-S.~Corasaniti\orcid{0000-0002-6386-7846}\inst{\ref{aff3},\ref{aff4}}
\and T.~Gayoux\orcid{0009-0008-9527-1490}\inst{\ref{aff3},\ref{aff5}}
\and G.~R\'acz\orcid{0000-0003-3906-5699}\inst{\ref{aff6}}
\and E.~Altamura\orcid{0000-0001-6973-1897}\inst{\ref{aff7},\ref{aff8}}
\and Z.~Sakr\orcid{0000-0002-4823-3757}\inst{\ref{aff9},\ref{aff10},\ref{aff11}}
\and C.~Carbone\orcid{0000-0003-0125-3563}\inst{\ref{aff12}}
\and M.~Baldi\orcid{0000-0003-4145-1943}\inst{\ref{aff13},\ref{aff14},\ref{aff15}}
\and C.~Giocoli\orcid{0000-0002-9590-7961}\inst{\ref{aff14},\ref{aff15}}
\and T.~Castro\orcid{0000-0002-6292-3228}\inst{\ref{aff2},\ref{aff16},\ref{aff17},\ref{aff18}}
\and F.~Pace\orcid{0000-0001-8039-0480}\inst{\ref{aff19},\ref{aff20},\ref{aff21},\ref{aff22}}
\and J.~E.~Taylor\orcid{0000-0002-6639-4183}\inst{\ref{aff23},\ref{aff24}}
\and S.~Borgani\orcid{0000-0001-6151-6439}\inst{\ref{aff25},\ref{aff17},\ref{aff2},\ref{aff16},\ref{aff18}}
\and O.~Luongo\orcid{0000-0001-7909-3577}\inst{\ref{aff26}}
\and C.~T.~Mpetha\orcid{0000-0002-7805-2500}\inst{\ref{aff27}}
\and R.~E.~Angulo\orcid{0000-0003-2953-3970}\inst{\ref{aff28},\ref{aff29}}
\and B.~Altieri\orcid{0000-0003-3936-0284}\inst{\ref{aff30}}
\and S.~Andreon\orcid{0000-0002-2041-8784}\inst{\ref{aff26}}
\and N.~Auricchio\orcid{0000-0003-4444-8651}\inst{\ref{aff14}}
\and C.~Baccigalupi\orcid{0000-0002-8211-1630}\inst{\ref{aff17},\ref{aff2},\ref{aff16},\ref{aff31}}
\and S.~Bardelli\orcid{0000-0002-8900-0298}\inst{\ref{aff14}}
\and P.~Battaglia\orcid{0000-0002-7337-5909}\inst{\ref{aff14}}
\and A.~Biviano\orcid{0000-0002-0857-0732}\inst{\ref{aff2},\ref{aff17}}
\and E.~Branchini\orcid{0000-0002-0808-6908}\inst{\ref{aff32},\ref{aff33},\ref{aff26}}
\and M.~Brescia\orcid{0000-0001-9506-5680}\inst{\ref{aff34},\ref{aff35}}
\and S.~Camera\orcid{0000-0003-3399-3574}\inst{\ref{aff19},\ref{aff20},\ref{aff21}}
\and V.~Capobianco\orcid{0000-0002-3309-7692}\inst{\ref{aff21}}
\and V.~F.~Cardone\inst{\ref{aff36},\ref{aff37}}
\and J.~Carretero\orcid{0000-0002-3130-0204}\inst{\ref{aff38},\ref{aff39}}
\and M.~Castellano\orcid{0000-0001-9875-8263}\inst{\ref{aff36}}
\and G.~Castignani\orcid{0000-0001-6831-0687}\inst{\ref{aff14}}
\and S.~Cavuoti\orcid{0000-0002-3787-4196}\inst{\ref{aff35},\ref{aff40}}
\and K.~C.~Chambers\orcid{0000-0001-6965-7789}\inst{\ref{aff41}}
\and A.~Cimatti\inst{\ref{aff42}}
\and C.~Colodro-Conde\inst{\ref{aff43}}
\and G.~Congedo\orcid{0000-0003-2508-0046}\inst{\ref{aff44}}
\and L.~Conversi\orcid{0000-0002-6710-8476}\inst{\ref{aff45},\ref{aff30}}
\and Y.~Copin\orcid{0000-0002-5317-7518}\inst{\ref{aff46}}
\and F.~Courbin\orcid{0000-0003-0758-6510}\inst{\ref{aff47},\ref{aff48},\ref{aff49}}
\and H.~M.~Courtois\orcid{0000-0003-0509-1776}\inst{\ref{aff50}}
\and M.~Cropper\orcid{0000-0003-4571-9468}\inst{\ref{aff51}}
\and H.~Degaudenzi\orcid{0000-0002-5887-6799}\inst{\ref{aff52}}
\and G.~De~Lucia\orcid{0000-0002-6220-9104}\inst{\ref{aff2}}
\and H.~Dole\orcid{0000-0002-9767-3839}\inst{\ref{aff53}}
\and F.~Dubath\orcid{0000-0002-6533-2810}\inst{\ref{aff52}}
\and X.~Dupac\inst{\ref{aff30}}
\and S.~Dusini\orcid{0000-0002-1128-0664}\inst{\ref{aff54}}
\and S.~Escoffier\orcid{0000-0002-2847-7498}\inst{\ref{aff55}}
\and M.~Farina\orcid{0000-0002-3089-7846}\inst{\ref{aff56}}
\and F.~Faustini\orcid{0000-0001-6274-5145}\inst{\ref{aff36}}
\and S.~Ferriol\inst{\ref{aff46}}
\and F.~Finelli\orcid{0000-0002-6694-3269}\inst{\ref{aff14},\ref{aff57}}
\and P.~Fosalba\orcid{0000-0002-1510-5214}\inst{\ref{aff58},\ref{aff59}}
\and S.~Fotopoulou\orcid{0000-0002-9686-254X}\inst{\ref{aff60}}
\and M.~Frailis\orcid{0000-0002-7400-2135}\inst{\ref{aff2}}
\and E.~Franceschi\orcid{0000-0002-0585-6591}\inst{\ref{aff14}}
\and M.~Fumana\orcid{0000-0001-6787-5950}\inst{\ref{aff12}}
\and L.~Gabarra\orcid{0000-0002-8486-8856}\inst{\ref{aff61}}
\and S.~Galeotta\orcid{0000-0002-3748-5115}\inst{\ref{aff2}}
\and K.~George\orcid{0000-0002-1734-8455}\inst{\ref{aff62}}
\and B.~Gillis\orcid{0000-0002-4478-1270}\inst{\ref{aff44}}
\and J.~Gracia-Carpio\orcid{0000-0003-4689-3134}\inst{\ref{aff63}}
\and A.~Grazian\orcid{0000-0002-5688-0663}\inst{\ref{aff64}}
\and F.~Grupp\inst{\ref{aff63},\ref{aff65}}
\and S.~V.~H.~Haugan\orcid{0000-0001-9648-7260}\inst{\ref{aff66}}
\and S.~Hemmati\orcid{0000-0003-2226-5395}\inst{\ref{aff67}}
\and W.~Holmes\orcid{0009-0007-8554-4646}\inst{\ref{aff68}}
\and F.~Hormuth\inst{\ref{aff69}}
\and A.~Hornstrup\orcid{0000-0002-3363-0936}\inst{\ref{aff70},\ref{aff71}}
\and K.~Jahnke\orcid{0000-0003-3804-2137}\inst{\ref{aff72}}
\and M.~Jhabvala\inst{\ref{aff27}}
\and B.~Joachimi\orcid{0000-0001-7494-1303}\inst{\ref{aff73}}
\and S.~Kermiche\orcid{0000-0002-0302-5735}\inst{\ref{aff55}}
\and A.~Kiessling\orcid{0000-0002-2590-1273}\inst{\ref{aff68}}
\and B.~Kubik\orcid{0009-0006-5823-4880}\inst{\ref{aff46}}
\and K.~Kuijken\orcid{0000-0002-3827-0175}\inst{\ref{aff74}}
\and M.~Kunz\orcid{0000-0002-3052-7394}\inst{\ref{aff75}}
\and H.~Kurki-Suonio\orcid{0000-0002-4618-3063}\inst{\ref{aff6},\ref{aff76}}
\and S.~Ligori\orcid{0000-0003-4172-4606}\inst{\ref{aff21}}
\and P.~B.~Lilje\orcid{0000-0003-4324-7794}\inst{\ref{aff66}}
\and V.~Lindholm\orcid{0000-0003-2317-5471}\inst{\ref{aff6},\ref{aff76}}
\and I.~Lloro\orcid{0000-0001-5966-1434}\inst{\ref{aff77}}
\and M.~Magliocchetti\orcid{0000-0001-9158-4838}\inst{\ref{aff56}}
\and G.~Mainetti\orcid{0000-0003-2384-2377}\inst{\ref{aff78}}
\and O.~Mansutti\orcid{0000-0001-5758-4658}\inst{\ref{aff2}}
\and O.~Marggraf\orcid{0000-0001-7242-3852}\inst{\ref{aff79}}
\and M.~Martinelli\orcid{0000-0002-6943-7732}\inst{\ref{aff36},\ref{aff37}}
\and N.~Martinet\orcid{0000-0003-2786-7790}\inst{\ref{aff80}}
\and F.~Marulli\orcid{0000-0002-8850-0303}\inst{\ref{aff81},\ref{aff14},\ref{aff15}}
\and R.~J.~Massey\orcid{0000-0002-6085-3780}\inst{\ref{aff82}}
\and N.~Mauri\orcid{0000-0001-8196-1548}\inst{\ref{aff42},\ref{aff15}}
\and S.~Maurogordato\inst{\ref{aff83}}
\and E.~Medinaceli\orcid{0000-0002-4040-7783}\inst{\ref{aff14}}
\and S.~Mei\orcid{0000-0002-2849-559X}\inst{\ref{aff84},\ref{aff85}}
\and M.~Meneghetti\orcid{0000-0003-1225-7084}\inst{\ref{aff14},\ref{aff15}}
\and E.~Merlin\orcid{0000-0001-6870-8900}\inst{\ref{aff64}}
\and G.~Meylan\orcid{0000-0001-6503-0209}\inst{\ref{aff86}}
\and P.~Monaco\orcid{0000-0003-2083-7564}\inst{\ref{aff25},\ref{aff2},\ref{aff16},\ref{aff17}}
\and A.~Mora\orcid{0000-0002-1922-8529}\inst{\ref{aff87}}
\and M.~Moresco\orcid{0000-0002-7616-7136}\inst{\ref{aff81},\ref{aff14}}
\and C.~Moretti\orcid{0000-0003-3314-8936}\inst{\ref{aff2},\ref{aff17},\ref{aff16}}
\and L.~Moscardini\orcid{0000-0002-3473-6716}\inst{\ref{aff81},\ref{aff14},\ref{aff15}}
\and E.~Munari\orcid{0000-0002-1751-5946}\inst{\ref{aff2},\ref{aff17}}
\and C.~Neissner\orcid{0000-0001-8524-4968}\inst{\ref{aff88},\ref{aff39}}
\and J.~W.~Nightingale\orcid{0000-0002-8987-7401}\inst{\ref{aff89}}
\and C.~Padilla\orcid{0000-0001-7951-0166}\inst{\ref{aff88}}
\and S.~Paltani\orcid{0000-0002-8108-9179}\inst{\ref{aff52}}
\and F.~Pasian\orcid{0000-0002-4869-3227}\inst{\ref{aff2}}
\and V.~Pettorino\orcid{0000-0002-4203-9320}\inst{\ref{aff90}}
\and A.~Pezzotta\orcid{0000-0003-0726-2268}\inst{\ref{aff26}}
\and S.~Pires\orcid{0000-0002-0249-2104}\inst{\ref{aff91}}
\and G.~Polenta\orcid{0000-0003-4067-9196}\inst{\ref{aff92}}
\and M.~Poncet\inst{\ref{aff93}}
\and L.~A.~Popa\inst{\ref{aff94}}
\and F.~Raison\orcid{0000-0002-7819-6918}\inst{\ref{aff63}}
\and A.~Renzi\orcid{0000-0001-9856-1970}\inst{\ref{aff95},\ref{aff54},\ref{aff14}}
\and J.~Rhodes\orcid{0000-0002-4485-8549}\inst{\ref{aff68}}
\and G.~Riccio\inst{\ref{aff35}}
\and I.~Risso\orcid{0000-0003-2525-7761}\inst{\ref{aff32},\ref{aff33},\ref{aff26}}
\and E.~Romelli\orcid{0000-0003-3069-9222}\inst{\ref{aff2}}
\and M.~Roncarelli\orcid{0000-0001-9587-7822}\inst{\ref{aff14}}
\and R.~Saglia\orcid{0000-0003-0378-7032}\inst{\ref{aff65},\ref{aff63}}
\and D.~Sapone\orcid{0000-0001-7089-4503}\inst{\ref{aff96}}
\and B.~Sartoris\orcid{0000-0003-1337-5269}\inst{\ref{aff65},\ref{aff2}}
\and P.~Schneider\orcid{0000-0001-8561-2679}\inst{\ref{aff79}}
\and T.~Schrabback\orcid{0000-0002-6987-7834}\inst{\ref{aff97}}
\and A.~Secroun\orcid{0000-0003-0505-3710}\inst{\ref{aff55}}
\and E.~Sefusatti\orcid{0000-0003-0473-1567}\inst{\ref{aff2},\ref{aff17},\ref{aff16}}
\and E.~Sihvola\orcid{0000-0003-1804-7715}\inst{\ref{aff98}}
\and P.~Simon\inst{\ref{aff79}}
\and C.~Sirignano\orcid{0000-0002-0995-7146}\inst{\ref{aff95},\ref{aff54}}
\and G.~Sirri\orcid{0000-0003-2626-2853}\inst{\ref{aff15}}
\and L.~Stanco\orcid{0000-0002-9706-5104}\inst{\ref{aff54}}
\and J.-L.~Starck\orcid{0000-0003-2177-7794}\inst{\ref{aff91}}
\and P.~Tallada-Cresp\'{i}\orcid{0000-0002-1336-8328}\inst{\ref{aff38},\ref{aff39}}
\and A.~N.~Taylor\inst{\ref{aff44}}
\and I.~Tereno\orcid{0000-0002-4537-6218}\inst{\ref{aff99},\ref{aff100}}
\and N.~Tessore\orcid{0000-0002-9696-7931}\inst{\ref{aff51}}
\and S.~Toft\orcid{0000-0003-3631-7176}\inst{\ref{aff101},\ref{aff102}}
\and R.~Toledo-Moreo\orcid{0000-0002-2997-4859}\inst{\ref{aff103},\ref{aff104}}
\and F.~Torradeflot\orcid{0000-0003-1160-1517}\inst{\ref{aff39},\ref{aff38}}
\and I.~Tutusaus\orcid{0000-0002-3199-0399}\inst{\ref{aff59},\ref{aff58},\ref{aff10}}
\and E.~A.~Valentijn\inst{\ref{aff105}}
\and J.~Valiviita\orcid{0000-0001-6225-3693}\inst{\ref{aff6},\ref{aff76}}
\and T.~Vassallo\orcid{0000-0001-6512-6358}\inst{\ref{aff2},\ref{aff62}}
\and Y.~Wang\orcid{0000-0002-4749-2984}\inst{\ref{aff67}}
\and J.~Weller\orcid{0000-0002-8282-2010}\inst{\ref{aff65},\ref{aff63}}
\and A.~Zacchei\orcid{0000-0003-0396-1192}\inst{\ref{aff2},\ref{aff17}}
\and G.~Zamorani\orcid{0000-0002-2318-301X}\inst{\ref{aff14}}
\and F.~M.~Zerbi\orcid{0000-0002-9996-973X}\inst{\ref{aff26}}
\and E.~Zucca\orcid{0000-0002-5845-8132}\inst{\ref{aff14}}
\and M.~Ballardini\orcid{0000-0003-4481-3559}\inst{\ref{aff106},\ref{aff107},\ref{aff14}}
\and P.~Bergamini\orcid{0000-0003-1383-9414}\inst{\ref{aff14}}
\and A.~Boucaud\orcid{0000-0001-7387-2633}\inst{\ref{aff84}}
\and E.~Bozzo\orcid{0000-0002-8201-1525}\inst{\ref{aff52}}
\and C.~Burigana\orcid{0000-0002-3005-5796}\inst{\ref{aff108},\ref{aff57}}
\and R.~Cabanac\orcid{0000-0001-6679-2600}\inst{\ref{aff10}}
\and M.~Calabrese\orcid{0000-0002-2637-2422}\inst{\ref{aff109},\ref{aff12}}
\and A.~Cappi\inst{\ref{aff83},\ref{aff14}}
\and F.~Caro\orcid{0009-0003-1053-0507}\inst{\ref{aff36}}
\and J.~A.~Escartin~Vigo\inst{\ref{aff63}}
\and J.~Garc\'ia-Bellido\orcid{0000-0002-9370-8360}\inst{\ref{aff9}}
\and T.~Gasparetto\orcid{0000-0002-7913-4866}\inst{\ref{aff36}}
\and J.~Macias-Perez\orcid{0000-0002-5385-2763}\inst{\ref{aff110}}
\and R.~Maoli\orcid{0000-0002-6065-3025}\inst{\ref{aff111},\ref{aff36}}
\and R.~B.~Metcalf\orcid{0000-0003-3167-2574}\inst{\ref{aff81},\ref{aff14}}
\and M.~P\"ontinen\orcid{0000-0001-5442-2530}\inst{\ref{aff6}}
\and E.~Sarpa\orcid{0000-0002-1256-655X}\inst{\ref{aff2}}
\and V.~Scottez\orcid{0009-0008-3864-940X}\inst{\ref{aff112},\ref{aff113}}
\and M.~Sereno\orcid{0000-0003-0302-0325}\inst{\ref{aff14},\ref{aff15}}
\and M.~Tenti\orcid{0000-0002-4254-5901}\inst{\ref{aff15}}
\and M.~Tucci\inst{\ref{aff52}}
\and M.~Viel\orcid{0000-0002-2642-5707}\inst{\ref{aff17},\ref{aff2},\ref{aff31},\ref{aff16},\ref{aff18}}
\and M.~Wiesmann\orcid{0009-0000-8199-5860}\inst{\ref{aff66}}
\and J.~A.~Acevedo~Barroso\orcid{0000-0002-9654-1711}\inst{\ref{aff68}}
\and Y.~Akrami\orcid{0000-0002-2407-7956}\inst{\ref{aff9},\ref{aff114}}
\and I.~T.~Andika\orcid{0000-0001-6102-9526}\inst{\ref{aff65}}
\and S.~Anselmi\orcid{0000-0002-3579-9583}\inst{\ref{aff54},\ref{aff95},\ref{aff115}}
\and M.~Archidiacono\orcid{0000-0003-4952-9012}\inst{\ref{aff116},\ref{aff117}}
\and G.~Aric\`o\orcid{0000-0002-2802-2928}\inst{\ref{aff15}}
\and F.~Atrio-Barandela\orcid{0000-0002-2130-2513}\inst{\ref{aff118}}
\and M.~Baes\orcid{0000-0002-3930-2757}\inst{\ref{aff119}}
\and L.~Bazzanini\orcid{0000-0003-0727-0137}\inst{\ref{aff106},\ref{aff14}}
\and D.~Bertacca\orcid{0000-0002-2490-7139}\inst{\ref{aff95},\ref{aff64},\ref{aff54}}
\and M.~Bethermin\orcid{0000-0002-3915-2015}\inst{\ref{aff120}}
\and F.~Beutler\orcid{0000-0003-0467-5438}\inst{\ref{aff44}}
\and A.~Blanchard\orcid{0000-0001-8555-9003}\inst{\ref{aff10}}
\and L.~Blot\orcid{0000-0002-9622-7167}\inst{\ref{aff121},\ref{aff3}}
\and H.~B\"ohringer\orcid{0000-0001-8241-4204}\inst{\ref{aff63},\ref{aff62},\ref{aff122}}
\and M.~L.~Brown\orcid{0000-0002-0370-8077}\inst{\ref{aff7}}
\and S.~Bruton\orcid{0000-0002-6503-5218}\inst{\ref{aff123}}
\and A.~Calabro\orcid{0000-0003-2536-1614}\inst{\ref{aff36}}
\and B.~Camacho~Quevedo\orcid{0000-0002-8789-4232}\inst{\ref{aff17},\ref{aff31},\ref{aff2}}
\and C.~S.~Carvalho\inst{\ref{aff100}}
\and F.~Cogato\orcid{0000-0003-4632-6113}\inst{\ref{aff81},\ref{aff14}}
\and T.~E.~Collett\orcid{0000-0001-5564-3140}\inst{\ref{aff124}}
\and S.~Conseil\orcid{0000-0002-3657-4191}\inst{\ref{aff46}}
\and A.~R.~Cooray\orcid{0000-0002-3892-0190}\inst{\ref{aff125}}
\and P.~Corcho-Caballero\orcid{0000-0001-6327-7080}\inst{\ref{aff105}}
\and M.~Costanzi\orcid{0000-0001-8158-1449}\inst{\ref{aff25},\ref{aff2},\ref{aff17}}
\and B.~Csizi\orcid{0000-0003-3227-6581}\inst{\ref{aff97}}
\and O.~Cucciati\orcid{0000-0002-9336-7551}\inst{\ref{aff14}}
\and T.~de~Boer\orcid{0000-0001-5486-2747}\inst{\ref{aff41}}
\and F.~De~Paolis\orcid{0000-0001-6460-7563}\inst{\ref{aff126},\ref{aff127},\ref{aff128}}
\and G.~Desprez\orcid{0000-0001-8325-1742}\inst{\ref{aff105}}
\and A.~D\'iaz-S\'anchez\orcid{0000-0003-0748-4768}\inst{\ref{aff129}}
\and S.~Di~Domizio\orcid{0000-0003-2863-5895}\inst{\ref{aff32},\ref{aff33}}
\and J.~M.~Diego\orcid{0000-0001-9065-3926}\inst{\ref{aff130}}
\and V.~Duret\orcid{0009-0009-0383-4960}\inst{\ref{aff55}}
\and Y.~Fang\orcid{0000-0002-0334-6950}\inst{\ref{aff65}}
\and A.~Farina\orcid{0009-0000-3420-929X}\inst{\ref{aff26},\ref{aff33}}
\and A.~Finoguenov\orcid{0000-0002-4606-5403}\inst{\ref{aff6}}
\and F.~Fontanot\orcid{0000-0003-4744-0188}\inst{\ref{aff2},\ref{aff17}}
\and A.~Franco\orcid{0000-0002-4761-366X}\inst{\ref{aff126},\ref{aff127},\ref{aff128}}
\and Y.~Fu\orcid{0000-0002-0759-0504}\inst{\ref{aff74},\ref{aff105}}
\and K.~Ganga\orcid{0000-0001-8159-8208}\inst{\ref{aff84}}
\and R.~Gavazzi\orcid{0000-0002-5540-6935}\inst{\ref{aff80},\ref{aff4}}
\and E.~Gaztanaga\orcid{0000-0001-9632-0815}\inst{\ref{aff59},\ref{aff58},\ref{aff124}}
\and Z.~Ghaffari\orcid{0000-0002-6467-8078}\inst{\ref{aff2},\ref{aff17}}
\and F.~Giacomini\orcid{0000-0002-3129-2814}\inst{\ref{aff15}}
\and F.~Gianotti\orcid{0000-0003-4666-119X}\inst{\ref{aff14}}
\and E.~J.~Gonzalez\orcid{0000-0002-0226-9893}\inst{\ref{aff59},\ref{aff131}}
\and G.~Gozaliasl\orcid{0000-0002-0236-919X}\inst{\ref{aff132},\ref{aff6}}
\and A.~Gruppuso\orcid{0000-0001-9272-5292}\inst{\ref{aff14},\ref{aff15}}
\and M.~Guidi\orcid{0000-0001-9408-1101}\inst{\ref{aff13},\ref{aff14}}
\and C.~M.~Gutierrez\orcid{0000-0001-7854-783X}\inst{\ref{aff43},\ref{aff133}}
\and A.~Hall\orcid{0000-0002-3139-8651}\inst{\ref{aff44}}
\and N.~A.~Hatch\orcid{0000-0001-5600-0534}\inst{\ref{aff134}}
\and C.~Hern\'andez-Monteagudo\orcid{0000-0001-5471-9166}\inst{\ref{aff133},\ref{aff43}}
\and H.~Hildebrandt\orcid{0000-0002-9814-3338}\inst{\ref{aff135}}
\and J.~Hjorth\orcid{0000-0002-4571-2306}\inst{\ref{aff136}}
\and J.~J.~E.~Kajava\orcid{0000-0002-3010-8333}\inst{\ref{aff137},\ref{aff138},\ref{aff139}}
\and Y.~Kang\orcid{0009-0000-8588-7250}\inst{\ref{aff52}}
\and V.~Kansal\orcid{0000-0002-4008-6078}\inst{\ref{aff140},\ref{aff141}}
\and D.~Karagiannis\orcid{0000-0002-4927-0816}\inst{\ref{aff106},\ref{aff142}}
\and J.~Kim\orcid{0000-0003-2776-2761}\inst{\ref{aff61}}
\and C.~C.~Kirkpatrick\inst{\ref{aff98}}
\and A.~Kov\'acs\orcid{0000-0002-5825-579X}\inst{\ref{aff143},\ref{aff144}}
\and I.~Kova{\v{c}}i{\'{c}}\orcid{0000-0001-6751-3263}\inst{\ref{aff119}}
\and K.~Koyama\orcid{0000-0001-6727-6915}\inst{\ref{aff124}}
\and S.~Kruk\orcid{0000-0001-8010-8879}\inst{\ref{aff30}}
\and M.~C.~Lam\orcid{0000-0002-9347-2298}\inst{\ref{aff44}}
\and M.~Lattanzi\orcid{0000-0003-1059-2532}\inst{\ref{aff107}}
\and L.~Legrand\orcid{0000-0003-0610-5252}\inst{\ref{aff145},\ref{aff53}}
\and M.~Lembo\orcid{0000-0002-5271-5070}\inst{\ref{aff4}}
\and G.~Leroy\orcid{0009-0004-2523-4425}\inst{\ref{aff146},\ref{aff82}}
\and G.~F.~Lesci\orcid{0000-0002-4607-2830}\inst{\ref{aff81},\ref{aff14}}
\and J.~Lesgourgues\orcid{0000-0001-7627-353X}\inst{\ref{aff147}}
\and L.~Linke\orcid{0000-0002-2622-8113}\inst{\ref{aff97}}
\and A.~Manj\'on-Garc\'ia\orcid{0000-0002-7413-8825}\inst{\ref{aff129}}
\and F.~Mannucci\orcid{0000-0002-4803-2381}\inst{\ref{aff148}}
\and F.~R.~Marleau\orcid{0000-0002-1442-2947}\inst{\ref{aff97}}
\and C.~J.~A.~P.~Martins\orcid{0000-0002-4886-9261}\inst{\ref{aff149},\ref{aff150}}
\and M.~Migliaccio\inst{\ref{aff151},\ref{aff152}}
\and M.~Miluzio\inst{\ref{aff30},\ref{aff153}}
\and G.~Morgante\inst{\ref{aff14}}
\and S.~Nadathur\orcid{0000-0001-9070-3102}\inst{\ref{aff124}}
\and K.~Naidoo\orcid{0000-0002-9182-1802}\inst{\ref{aff124},\ref{aff72}}
\and A.~Navarro-Alsina\orcid{0000-0002-3173-2592}\inst{\ref{aff79}}
\and S.~Nesseris\orcid{0000-0002-0567-0324}\inst{\ref{aff9}}
\and F.~Oppizzi\orcid{0000-0003-3904-8370}\inst{\ref{aff33},\ref{aff54},\ref{aff95}}
\and D.~Paoletti\orcid{0000-0003-4761-6147}\inst{\ref{aff14},\ref{aff57}}
\and F.~Passalacqua\orcid{0000-0002-8606-4093}\inst{\ref{aff54}}
\and K.~Paterson\orcid{0000-0001-8340-3486}\inst{\ref{aff72}}
\and L.~Patrizii\inst{\ref{aff15}}
\and C.~Pattison\orcid{0000-0003-3272-2617}\inst{\ref{aff124}}
\and R.~Paviot\orcid{0009-0002-8108-3460}\inst{\ref{aff91},\ref{aff93}}
\and A.~Pisani\orcid{0000-0002-6146-4437}\inst{\ref{aff55}}
\and D.~Potter\orcid{0000-0002-0757-5195}\inst{\ref{aff154}}
\and G.~W.~Pratt\inst{\ref{aff91}}
\and S.~Quai\orcid{0000-0002-0449-8163}\inst{\ref{aff81},\ref{aff14}}
\and M.~Radovich\orcid{0000-0002-3585-866X}\inst{\ref{aff64}}
\and G.~Rodighiero\orcid{0000-0002-9415-2296}\inst{\ref{aff95},\ref{aff64}}
\and W.~Roster\orcid{0000-0002-9149-6528}\inst{\ref{aff63}}
\and S.~Sacquegna\orcid{0000-0002-8433-6630}\inst{\ref{aff155}}
\and M.~Sahl\'en\orcid{0000-0003-0973-4804}\inst{\ref{aff156}}
\and D.~B.~Sanders\orcid{0000-0002-1233-9998}\inst{\ref{aff41}}
\and A.~Schneider\orcid{0000-0001-7055-8104}\inst{\ref{aff154}}
\and D.~Sciotti\orcid{0009-0008-4519-2620}\inst{\ref{aff36},\ref{aff37}}
\and E.~Sellentin\orcid{0009-0002-2655-3458}\inst{\ref{aff157},\ref{aff74}}
\and S.~Serjeant\orcid{0000-0002-0517-7943}\inst{\ref{aff158}}
\and F.~Shankar\orcid{0000-0001-8973-5051}\inst{\ref{aff159}}
\and L.~C.~Smith\orcid{0000-0002-3259-2771}\inst{\ref{aff160}}
\and J.~G.~Sorce\orcid{0000-0002-2307-2432}\inst{\ref{aff161},\ref{aff53}}
\and I.~Szapudi\orcid{0000-0003-2274-0301}\inst{\ref{aff41}}
\and K.~Tanidis\orcid{0000-0001-9843-5130}\inst{\ref{aff162}}
\and C.~Tao\orcid{0000-0001-7961-8177}\inst{\ref{aff55}}
\and F.~Tarsitano\orcid{0000-0002-5919-0238}\inst{\ref{aff163},\ref{aff164},\ref{aff52}}
\and G.~Testera\orcid{0000-0003-2970-766X}\inst{\ref{aff33}}
\and R.~Teyssier\orcid{0000-0001-7689-0933}\inst{\ref{aff165}}
\and S.~Tosi\orcid{0000-0002-7275-9193}\inst{\ref{aff32},\ref{aff26},\ref{aff33}}
\and A.~Troja\orcid{0000-0003-0239-4595}\inst{\ref{aff2}}
\and C.~Uhlemann\orcid{0000-0001-7831-1579}\inst{\ref{aff166},\ref{aff89}}
\and C.~Valieri\inst{\ref{aff15}}
\and A.~Venhola\orcid{0000-0001-6071-4564}\inst{\ref{aff167}}
\and D.~Vergani\orcid{0000-0003-0898-2216}\inst{\ref{aff14}}
\and G.~Verza\orcid{0000-0002-1886-8348}\thanks{Deceased}\inst{\ref{aff168},\ref{aff169}}
\and P.~Vielzeuf\orcid{0000-0003-2035-9339}\inst{\ref{aff55}}
\and S.~Vinciguerra\orcid{0009-0005-4018-3184}\inst{\ref{aff80}}
\and M.~von~Wietersheim-Kramsta\orcid{0000-0003-4986-5091}\inst{\ref{aff82},\ref{aff146}}
\and L.~Wang\orcid{0000-0002-6736-9158}\inst{\ref{aff170},\ref{aff105}}
\and A.~H.~Wright\orcid{0000-0001-7363-7932}\inst{\ref{aff135}}
\and H.~W.~Yeung\orcid{0000-0002-4993-9014}\inst{\ref{aff44}}}
										   
%%%% please do not edit the affiliation list -- contact ECEB Bureau for changes
\institute{Dipartimento di Fisica ``G. Occhialini", Universit\`a degli Studi di Milano Bicocca, Piazza della Scienza 3, 20126 Milano, Italy\label{aff1}
\and
INAF-Osservatorio Astronomico di Trieste, Via G. B. Tiepolo 11, 34143 Trieste, Italy\label{aff2}
\and
Laboratoire d'etude de l'Univers et des phenomenes eXtremes, Observatoire de Paris, Universit\'e PSL, Sorbonne Universit\'e, CNRS, 92190 Meudon, France\label{aff3}
\and
Institut d'Astrophysique de Paris, UMR 7095, CNRS, and Sorbonne Universit\'e, 98 bis boulevard Arago, 75014 Paris, France\label{aff4}
\and
Universit\'e Paris-Cit\'e, 5 Rue Thomas Mann, 75013, Paris, France\label{aff5}
\and
Department of Physics, P.O. Box 64, University of Helsinki, 00014 Helsinki, Finland\label{aff6}
\and
Jodrell Bank Centre for Astrophysics, Department of Physics and Astronomy, University of Manchester, Oxford Road, Manchester M13 9PL, UK\label{aff7}
\and
Yusuf Hamied Department of Chemistry, University of Cambridge, Lensfield Road, Cambridge CB2 1EW, UK\label{aff8}
\and
Instituto de F\'isica Te\'orica UAM-CSIC, Campus de Cantoblanco, 28049 Madrid, Spain\label{aff9}
\and
Institut de Recherche en Astrophysique et Plan\'etologie (IRAP), Universit\'e de Toulouse, CNRS, UPS, CNES, 14 Av. Edouard Belin, 31400 Toulouse, France\label{aff10}
\and
Universit\'e St Joseph; Faculty of Sciences, Beirut, Lebanon\label{aff11}
\and
INAF-IASF Milano, Via Alfonso Corti 12, 20133 Milano, Italy\label{aff12}
\and
Dipartimento di Fisica e Astronomia, Universit\`a di Bologna, Via Gobetti 93/2, 40129 Bologna, Italy\label{aff13}
\and
INAF-Osservatorio di Astrofisica e Scienza dello Spazio di Bologna, Via Piero Gobetti 93/3, 40129 Bologna, Italy\label{aff14}
\and
INFN-Sezione di Bologna, Viale Berti Pichat 6/2, 40127 Bologna, Italy\label{aff15}
\and
INFN, Sezione di Trieste, Via Valerio 2, 34127 Trieste TS, Italy\label{aff16}
\and
IFPU, Institute for Fundamental Physics of the Universe, via Beirut 2, 34151 Trieste, Italy\label{aff17}
\and
ICSC - Centro Nazionale di Ricerca in High Performance Computing, Big Data e Quantum Computing, Via Magnanelli 2, Bologna, Italy\label{aff18}
\and
Dipartimento di Fisica, Universit\`a degli Studi di Torino, Via P. Giuria 1, 10125 Torino, Italy\label{aff19}
\and
INFN-Sezione di Torino, Via P. Giuria 1, 10125 Torino, Italy\label{aff20}
\and
INAF-Osservatorio Astrofisico di Torino, Via Osservatorio 20, 10025 Pino Torinese (TO), Italy\label{aff21}
\and
Instituto de Astrof\'isica e Ci\^encias do Espa\c{c}o, Faculdade de Ci\^encias, Universidade de Lisboa, Campo Grande, 1749-016 Lisboa, Portugal\label{aff22}
\and
Department of Physics and Astronomy, University of Waterloo, Waterloo, Ontario N2L 3G1, Canada\label{aff23}
\and
Waterloo Centre for Astrophysics, University of Waterloo, Waterloo, Ontario N2L 3G1, Canada\label{aff24}
\and
Dipartimento di Fisica - Sezione di Astronomia, Universit\`a di Trieste, Via Tiepolo 11, 34131 Trieste, Italy\label{aff25}
\and
INAF-Osservatorio Astronomico di Brera, Via Brera 28, 20122 Milano, Italy\label{aff26}
\and
NASA Goddard Space Flight Center, Greenbelt, MD 20771, USA\label{aff27}
\and
Donostia International Physics Center (DIPC), Paseo Manuel de Lardizabal, 4, 20018, Donostia-San Sebasti\'an, Guipuzkoa, Spain\label{aff28}
\and
IKERBASQUE, Basque Foundation for Science, 48013, Bilbao, Spain\label{aff29}
\and
ESAC/ESA, Camino Bajo del Castillo, s/n., Urb. Villafranca del Castillo, 28692 Villanueva de la Ca\~nada, Madrid, Spain\label{aff30}
\and
SISSA, International School for Advanced Studies, Via Bonomea 265, 34136 Trieste TS, Italy\label{aff31}
\and
Dipartimento di Fisica, Universit\`a di Genova, Via Dodecaneso 33, 16146, Genova, Italy\label{aff32}
\and
INFN-Sezione di Genova, Via Dodecaneso 33, 16146, Genova, Italy\label{aff33}
\and
Department of Physics "E. Pancini", University Federico II, Via Cinthia 6, 80126, Napoli, Italy\label{aff34}
\and
INAF-Osservatorio Astronomico di Capodimonte, Via Moiariello 16, 80131 Napoli, Italy\label{aff35}
\and
INAF-Osservatorio Astronomico di Roma, Via Frascati 33, 00078 Monteporzio Catone, Italy\label{aff36}
\and
INFN-Sezione di Roma, Piazzale Aldo Moro, 2 - c/o Dipartimento di Fisica, Edificio G. Marconi, 00185 Roma, Italy\label{aff37}
\and
Centro de Investigaciones Energ\'eticas, Medioambientales y Tecnol\'ogicas (CIEMAT), Avenida Complutense 40, 28040 Madrid, Spain\label{aff38}
\and
Port d'Informaci\'{o} Cient\'{i}fica, Campus UAB, C. Albareda s/n, 08193 Bellaterra (Barcelona), Spain\label{aff39}
\and
INFN -- Sezione di Napoli, Via Cinthia 6, 80126, Napoli, Italy\label{aff40}
\and
Institute for Astronomy, University of Hawaii, 2680 Woodlawn Drive, Honolulu, HI 96822, USA\label{aff41}
\and
Dipartimento di Fisica e Astronomia "Augusto Righi" - Alma Mater Studiorum Universit\`a di Bologna, Viale Berti Pichat 6/2, 40127 Bologna, Italy\label{aff42}
\and
Instituto de Astrof\'{\i}sica de Canarias, E-38205 La Laguna, Tenerife, Spain\label{aff43}
\and
Institute for Astronomy, University of Edinburgh, Royal Observatory, Blackford Hill, Edinburgh EH9 3HJ, UK\label{aff44}
\and
European Space Agency/ESRIN, Largo Galileo Galilei 1, 00044 Frascati, Roma, Italy\label{aff45}
\and
Universit\'e Claude Bernard Lyon 1, CNRS/IN2P3, IP2I Lyon, UMR 5822, Villeurbanne, F-69100, France\label{aff46}
\and
Institut de Ci\`{e}ncies del Cosmos (ICCUB), Universitat de Barcelona (IEEC-UB), Mart\'{i} i Franqu\`{e}s 1, 08028 Barcelona, Spain\label{aff47}
\and
Instituci\'o Catalana de Recerca i Estudis Avan\c{c}ats (ICREA), Passeig de Llu\'{\i}s Companys 23, 08010 Barcelona, Spain\label{aff48}
\and
Institut de Ciencies de l'Espai (IEEC-CSIC), Campus UAB, Carrer de Can Magrans, s/n Cerdanyola del Vall\'es, 08193 Barcelona, Spain\label{aff49}
\and
UCB Lyon 1, CNRS/IN2P3, IUF, IP2I Lyon, 4 rue Enrico Fermi, 69622 Villeurbanne, France\label{aff50}
\and
Mullard Space Science Laboratory, University College London, Holmbury St Mary, Dorking, Surrey RH5 6NT, UK\label{aff51}
\and
Department of Astronomy, University of Geneva, ch. d'Ecogia 16, 1290 Versoix, Switzerland\label{aff52}
\and
Universit\'e Paris-Saclay, CNRS, Institut d'astrophysique spatiale, 91405, Orsay, France\label{aff53}
\and
INFN-Padova, Via Marzolo 8, 35131 Padova, Italy\label{aff54}
\and
Aix-Marseille Universit\'e, CNRS/IN2P3, CPPM, Marseille, France\label{aff55}
\and
INAF-Istituto di Astrofisica e Planetologia Spaziali, via del Fosso del Cavaliere, 100, 00100 Roma, Italy\label{aff56}
\and
INFN-Bologna, Via Irnerio 46, 40126 Bologna, Italy\label{aff57}
\and
Institut d'Estudis Espacials de Catalunya (IEEC),  Edifici RDIT, Campus UPC, 08860 Castelldefels, Barcelona, Spain\label{aff58}
\and
Institute of Space Sciences (ICE, CSIC), Campus UAB, Carrer de Can Magrans, s/n, 08193 Barcelona, Spain\label{aff59}
\and
School of Physics, HH Wills Physics Laboratory, University of Bristol, Tyndall Avenue, Bristol, BS8 1TL, UK\label{aff60}
\and
Department of Physics, University of Oxford, Keble Road, Oxford OX1 3RH, UK\label{aff61}
\and
University Observatory, LMU Faculty of Physics, Scheinerstr.~1, 81679 Munich, Germany\label{aff62}
\and
Max Planck Institute for Extraterrestrial Physics, Giessenbachstr. 1, 85748 Garching, Germany\label{aff63}
\and
INAF-Osservatorio Astronomico di Padova, Via dell'Osservatorio 5, 35122 Padova, Italy\label{aff64}
\and
Universit\"ats-Sternwarte M\"unchen, Fakult\"at f\"ur Physik, Ludwig-Maximilians-Universit\"at M\"unchen, Scheinerstr.~1, 81679 M\"unchen, Germany\label{aff65}
\and
Institute of Theoretical Astrophysics, University of Oslo, P.O. Box 1029 Blindern, 0315 Oslo, Norway\label{aff66}
\and
Caltech/IPAC, 1200 E. California Blvd., Pasadena, CA 91125, USA\label{aff67}
\and
Jet Propulsion Laboratory, California Institute of Technology, 4800 Oak Grove Drive, Pasadena, CA, 91109, USA\label{aff68}
\and
Felix Hormuth Engineering, Goethestr. 17, 69181 Leimen, Germany\label{aff69}
\and
Technical University of Denmark, Elektrovej 327, 2800 Kgs. Lyngby, Denmark\label{aff70}
\and
Cosmic Dawn Center (DAWN), Denmark\label{aff71}
\and
Max-Planck-Institut f\"ur Astronomie, K\"onigstuhl 17, 69117 Heidelberg, Germany\label{aff72}
\and
Department of Physics and Astronomy, University College London, Gower Street, London WC1E 6BT, UK\label{aff73}
\and
Leiden Observatory, Leiden University, Einsteinweg 55, 2333 CC Leiden, The Netherlands\label{aff74}
\and
Universit\'e de Gen\`eve, D\'epartement de Physique Th\'eorique and Centre for Astroparticle Physics, 24 quai Ernest-Ansermet, CH-1211 Gen\`eve 4, Switzerland\label{aff75}
\and
Helsinki Institute of Physics, Gustaf H{\"a}llstr{\"o}min katu 2, University of Helsinki, 00014 Helsinki, Finland\label{aff76}
\and
SKAO, Jodrell Bank, Lower Withington, Macclesfield SK11 9FT, UK\label{aff77}
\and
Centre de Calcul de l'IN2P3/CNRS, 21 avenue Pierre de Coubertin 69627 Villeurbanne Cedex, France\label{aff78}
\and
Universit\"at Bonn, Argelander-Institut f\"ur Astronomie, Auf dem H\"ugel 71, 53121 Bonn, Germany\label{aff79}
\and
Aix-Marseille Universit\'e, CNRS, CNES, LAM, Marseille, France\label{aff80}
\and
Dipartimento di Fisica e Astronomia "Augusto Righi" - Alma Mater Studiorum Universit\`a di Bologna, via Piero Gobetti 93/2, 40129 Bologna, Italy\label{aff81}
\and
Department of Physics, Institute for Computational Cosmology, Durham University, South Road, Durham, DH1 3LE, UK\label{aff82}
\and
Universit\'e C\^{o}te d'Azur, Observatoire de la C\^{o}te d'Azur, CNRS, Laboratoire Lagrange, Bd de l'Observatoire, CS 34229, 06304 Nice cedex 4, France\label{aff83}
\and
Universit\'e Paris Cit\'e, CNRS, Astroparticule et Cosmologie, 75013 Paris, France\label{aff84}
\and
CNRS-UCB International Research Laboratory, Centre Pierre Bin\'etruy, IRL2007, CPB-IN2P3, Berkeley, USA\label{aff85}
\and
Institute of Physics, Laboratory of Astrophysics, Ecole Polytechnique F\'ed\'erale de Lausanne (EPFL), Observatoire de Sauverny, 1290 Versoix, Switzerland\label{aff86}
\and
Telespazio UK S.L. for European Space Agency (ESA), Camino bajo del Castillo, s/n, Urbanizacion Villafranca del Castillo, Villanueva de la Ca\~nada, 28692 Madrid, Spain\label{aff87}
\and
Institut de F\'{i}sica d'Altes Energies (IFAE), The Barcelona Institute of Science and Technology, Campus UAB, 08193 Bellaterra (Barcelona), Spain\label{aff88}
\and
School of Mathematics, Statistics and Physics, Newcastle University, Herschel Building, Newcastle-upon-Tyne, NE1 7RU, UK\label{aff89}
\and
European Space Agency/ESTEC, Keplerlaan 1, 2201 AZ Noordwijk, The Netherlands\label{aff90}
\and
Universit\'e Paris-Saclay, Universit\'e Paris Cit\'e, CEA, CNRS, AIM, 91191, Gif-sur-Yvette, France\label{aff91}
\and
Space Science Data Center, Italian Space Agency, via del Politecnico snc, 00133 Roma, Italy\label{aff92}
\and
Centre National d'Etudes Spatiales -- Centre spatial de Toulouse, 18 avenue Edouard Belin, 31401 Toulouse Cedex 9, France\label{aff93}
\and
Institute of Space Science, Str. Atomistilor, nr. 409 M\u{a}gurele, Ilfov, 077125, Romania\label{aff94}
\and
Dipartimento di Fisica e Astronomia "G. Galilei", Universit\`a di Padova, Via Marzolo 8, 35131 Padova, Italy\label{aff95}
\and
Departamento de F\'isica, FCFM, Universidad de Chile, Blanco Encalada 2008, Santiago, Chile\label{aff96}
\and
Universit\"at Innsbruck, Institut f\"ur Astro- und Teilchenphysik, Technikerstr. 25/8, 6020 Innsbruck, Austria\label{aff97}
\and
Department of Physics and Helsinki Institute of Physics, Gustaf H\"allstr\"omin katu 2, University of Helsinki, 00014 Helsinki, Finland\label{aff98}
\and
Departamento de F\'isica, Faculdade de Ci\^encias, Universidade de Lisboa, Edif\'icio C8, Campo Grande, PT1749-016 Lisboa, Portugal\label{aff99}
\and
Instituto de Astrof\'isica e Ci\^encias do Espa\c{c}o, Faculdade de Ci\^encias, Universidade de Lisboa, Tapada da Ajuda, 1349-018 Lisboa, Portugal\label{aff100}
\and
Cosmic Dawn Center (DAWN)\label{aff101}
\and
Niels Bohr Institute, University of Copenhagen, Jagtvej 128, 2200 Copenhagen, Denmark\label{aff102}
\and
Universidad Polit\'ecnica de Cartagena, Departamento de Electr\'onica y Tecnolog\'ia de Computadoras,  Plaza del Hospital 1, 30202 Cartagena, Spain\label{aff103}
\and
European University of Technology EUt+, European Union\label{aff104}
\and
Kapteyn Astronomical Institute, University of Groningen, PO Box 800, 9700 AV Groningen, The Netherlands\label{aff105}
\and
Dipartimento di Fisica e Scienze della Terra, Universit\`a degli Studi di Ferrara, Via Giuseppe Saragat 1, 44122 Ferrara, Italy\label{aff106}
\and
Istituto Nazionale di Fisica Nucleare, Sezione di Ferrara, Via Giuseppe Saragat 1, 44122 Ferrara, Italy\label{aff107}
\and
INAF, Istituto di Radioastronomia, Via Piero Gobetti 101, 40129 Bologna, Italy\label{aff108}
\and
Astronomical Observatory of the Autonomous Region of the Aosta Valley (OAVdA), Loc. Lignan 39, I-11020, Nus (Aosta Valley), Italy\label{aff109}
\and
Univ. Grenoble Alpes, CNRS, Grenoble INP, LPSC-IN2P3, 53, Avenue des Martyrs, 38000, Grenoble, France\label{aff110}
\and
Dipartimento di Fisica, Sapienza Universit\`a di Roma, Piazzale Aldo Moro 2, 00185 Roma, Italy\label{aff111}
\and
Institut d'Astrophysique de Paris, 98bis Boulevard Arago, 75014, Paris, France\label{aff112}
\and
ICL, Junia, Universit\'e Catholique de Lille, LITL, 59000 Lille, France\label{aff113}
\and
CERCA/ISO, Department of Physics, Case Western Reserve University, 10900 Euclid Avenue, Cleveland, OH 44106, USA\label{aff114}
\and
Laboratoire Univers et Th\'eorie, Observatoire de Paris, Universit\'e PSL, Universit\'e Paris Cit\'e, CNRS, 92190 Meudon, France\label{aff115}
\and
Dipartimento di Fisica "Aldo Pontremoli", Universit\`a degli Studi di Milano, Via Celoria 16, 20133 Milano, Italy\label{aff116}
\and
INFN-Sezione di Milano, Via Celoria 16, 20133 Milano, Italy\label{aff117}
\and
Departamento de F{\'\i}sica Fundamental. Universidad de Salamanca. Plaza de la Merced s/n. 37008 Salamanca, Spain\label{aff118}
\and
Universiteit Gent, Department of Physics and Astronomy, Proeftuinstraat 86 N3, 9000 Ghent, Belgium
\label{aff119}
\and
Universit\'e de Strasbourg, CNRS, Observatoire astronomique de Strasbourg, UMR 7550, 67000 Strasbourg, France\label{aff120}
\and
Center for Data-Driven Discovery, Kavli IPMU (WPI), UTIAS, The University of Tokyo, Kashiwa, Chiba 277-8583, Japan\label{aff121}
\and
Max-Planck-Institut f\"ur Physik, Boltzmannstr. 8, 85748 Garching, Germany\label{aff122}
\and
California Institute of Technology, 1200 E California Blvd, Pasadena, CA 91125, USA\label{aff123}
\and
Institute of Cosmology and Gravitation, University of Portsmouth, Portsmouth PO1 3FX, UK\label{aff124}
\and
Department of Physics \& Astronomy, University of California Irvine, Irvine CA 92697, USA\label{aff125}
\and
Department of Mathematics and Physics E. De Giorgi, University of Salento, Via per Arnesano, CP-I93, 73100, Lecce, Italy\label{aff126}
\and
INFN, Sezione di Lecce, Via per Arnesano, CP-193, 73100, Lecce, Italy\label{aff127}
\and
INAF-Sezione di Lecce, c/o Dipartimento Matematica e Fisica, Via per Arnesano, 73100, Lecce, Italy\label{aff128}
\and
Departamento F\'isica Aplicada, Universidad Polit\'ecnica de Cartagena, Campus Muralla del Mar, 30202 Cartagena, Murcia, Spain\label{aff129}
\and
Instituto de F\'isica de Cantabria, Edificio Juan Jord\'a, Avenida de los Castros, 39005 Santander, Spain\label{aff130}
\and
Instituto de Astronomia Teorica y Experimental (IATE-CONICET), Laprida 854, X5000BGR, C\'ordoba, Argentina\label{aff131}
\and
Department of Computer Science, Aalto University, PO Box 15400, Espoo, FI-00 076, Finland\label{aff132}
\and
Universidad de La Laguna, Dpto. Astrof\'\i sica, E-38206 La Laguna, Tenerife, Spain\label{aff133}
\and
School of Physics and Astronomy, University of Nottingham, University Park, Nottingham NG7 2RD, UK\label{aff134}
\and
Ruhr University Bochum, Faculty of Physics and Astronomy, Astronomical Institute (AIRUB), German Centre for Cosmological Lensing (GCCL), 44780 Bochum, Germany\label{aff135}
\and
DARK, Niels Bohr Institute, University of Copenhagen, Jagtvej 155, 2200 Copenhagen, Denmark\label{aff136}
\and
Department of Physics and Astronomy, Vesilinnantie 5, University of Turku, 20014 Turku, Finland\label{aff137}
\and
Finnish Centre for Astronomy with ESO (FINCA), Quantum, Vesilinnantie 5, University of Turku, 20014 Turku, Finland\label{aff138}
\and
Serco for European Space Agency (ESA), Camino bajo del Castillo, s/n, Urbanizacion Villafranca del Castillo, Villanueva de la Ca\~nada, 28692 Madrid, Spain\label{aff139}
\and
ARC Centre of Excellence for Dark Matter Particle Physics, Melbourne, Australia\label{aff140}
\and
Centre for Astrophysics \& Supercomputing, Swinburne University of Technology,  Hawthorn, Victoria 3122, Australia\label{aff141}
\and
Department of Physics and Astronomy, University of the Western Cape, Bellville, Cape Town, 7535, South Africa\label{aff142}
\and
MTA-CSFK Lend\"ulet Large-Scale Structure Research Group, Konkoly-Thege Mikl\'os \'ut 15-17, H-1121 Budapest, Hungary\label{aff143}
\and
Konkoly Observatory, HUN-REN CSFK, MTA Centre of Excellence, Budapest, Konkoly Thege Mikl\'os {\'u}t 15-17. H-1121, Hungary\label{aff144}
\and
Brazilian Center for Research in Physics (CBPF), Dr. Xavier Sigaud st. 150, zip 22290-180, Rio de Janeiro, RJ, Brazil\label{aff145}
\and
Department of Physics, Centre for Extragalactic Astronomy, Durham University, South Road, Durham, DH1 3LE, UK\label{aff146}
\and
Institute for Theoretical Particle Physics and Cosmology (TTK), RWTH Aachen University, 52056 Aachen, Germany\label{aff147}
\and
INAF-Osservatorio Astrofisico di Arcetri, Largo E. Fermi 5, 50125, Firenze, Italy\label{aff148}
\and
Centro de Astrof\'{\i}sica da Universidade do Porto, Rua das Estrelas, 4150-762 Porto, Portugal\label{aff149}
\and
Instituto de Astrof\'isica e Ci\^encias do Espa\c{c}o, Universidade do Porto, CAUP, Rua das Estrelas, PT4150-762 Porto, Portugal\label{aff150}
\and
Dipartimento di Fisica, Universit\`a di Roma Tor Vergata, Via della Ricerca Scientifica 1, Roma, Italy\label{aff151}
\and
INFN, Sezione di Roma 2, Via della Ricerca Scientifica 1, Roma, Italy\label{aff152}
\and
HE Space for European Space Agency (ESA), Camino bajo del Castillo, s/n, Urbanizacion Villafranca del Castillo, Villanueva de la Ca\~nada, 28692 Madrid, Spain\label{aff153}
\and
Department of Astrophysics, University of Zurich, Winterthurerstrasse 190, 8057 Zurich, Switzerland\label{aff154}
\and
INAF - Osservatorio Astronomico d'Abruzzo, Via Maggini, 64100, Teramo, Italy\label{aff155}
\and
Theoretical astrophysics, Department of Physics and Astronomy, Uppsala University, Box 516, 751 37 Uppsala, Sweden\label{aff156}
\and
Mathematical Institute, University of Leiden, Einsteinweg 55, 2333 CA Leiden, The Netherlands\label{aff157}
\and
School of Physical Sciences, The Open University, Milton Keynes, MK7 6AA, UK\label{aff158}
\and
School of Physics \& Astronomy, University of Southampton, Highfield Campus, Southampton SO17 1BJ, UK\label{aff159}
\and
Institute of Astronomy, University of Cambridge, Madingley Road, Cambridge CB3 0HA, UK\label{aff160}
\and
Univ. Lille, CNRS, Centrale Lille, UMR 9189 CRIStAL, 59000 Lille, France\label{aff161}
\and
Center for Astrophysics and Cosmology, University of Nova Gorica, Nova Gorica, Slovenia\label{aff162}
\and
Kobayashi-Maskawa Institute for the Origin of Particles and the Universe, Nagoya University, Chikusa-ku, Nagoya, 464-8602, Japan\label{aff163}
\and
Institute for Particle Physics and Astrophysics, Dept. of Physics, ETH Zurich, Wolfgang-Pauli-Strasse 27, 8093 Zurich, Switzerland\label{aff164}
\and
Department of Astrophysical Sciences, Peyton Hall, Princeton University, Princeton, NJ 08544, USA\label{aff165}
\and
Fakult\"at f\"ur Physik, Universit\"at Bielefeld, Postfach 100131, 33501 Bielefeld, Germany\label{aff166}
\and
Space physics and astronomy research unit, University of Oulu, Pentti Kaiteran katu 1, FI-90014 Oulu, Finland\label{aff167}
\and
International Centre for Theoretical Physics (ICTP), Strada Costiera 11, 34151 Trieste, Italy\label{aff168}
\and
Center for Computational Astrophysics, Flatiron Institute, 162 5th Avenue, 10010, New York, NY, USA\label{aff169}
\and
SRON Netherlands Institute for Space Research, Landleven 12, 9747 AD, Groningen, The Netherlands\label{aff170}}     % Contains the author and affiliation list;
                   % This list can be obtained from the Publication Portal

 %\date{\bf *Version 3.25*, 16 Jan. 2026}

% 
% Put your abstract here
% -- through a newly developed pipeline -- 

\abstract{We study the shape of three-dimensional and projected dark-matter halo profiles extracted from cosmological $N$-body simulations in $\Lambda$CDM and non-standard cosmologies, using the \DUSTGRAINPATHFINDER and \DEMNUni suites. The models considered include massive neutrinos, $f(\mathcal{R})$ gravity, and dynamical dark energy. By comparing density, mass, velocity-dispersion, and excess-surface-density profiles up to $5\,r_{500{\rm c}}$, we quantify the differential imprint of non-standard physics on halo structure in view of \Euclid cluster WL studies. Our main analysis is performed at $z=1.1$, a high-redshift regime where the weak-lensing signal-to-noise starts to degrade, providing a conservative stress test for detectability; for \DUSTGRAINPATHFINDER we additionally analyse $z=0.5$ and $z=0.3$ snapshots. In low-mass haloes ($M_{\rm 200c}<7\times10^{13}\,M_\odot$), $f(\mathcal{R})$ gravity produces deviations of order $10\,\%$ in projected and three-dimensional profiles, especially in the outskirts where screening is less efficient. Massive neutrinos partially reduce this signal, reflecting the competition between free streaming and fifth-force-enhanced growth. Dynamical dark energy and massive-neutrino cosmologies generally induce smaller, few-percent deviations, with the largest effects again found in low-mass haloes and at large radii. Under simplified assumptions for \Euclid WL, detecting such profile differences at $z=1.1$ requires stacks of $\sim10^5$ haloes, while a few thousands haloes may be sufficient at $z\lesssim0.5$. This further calls for the need of integrating such precise modelling of non-standard effects -- along with other observational effects -- in any likelihood involving \Euclid WL masses to avoid non-negligible systematic biases. Concentration--mass relations show weaker cosmology dependence, typically at the $\sim5\,\%$ level. Although full forecasts require the inclusion of baryonic physics and observational systematics, our dark-matter-only analysis provides a controlled baseline for identifying where non-standard cosmological signatures are expected to be largest and most relevant for upcoming \Euclid data.}

%
% Add short versions of title and author list for page headings
%
   \titlerunning{Halo profiles in $\Lambda$CDM and non-standard cosmologies}
   \authorrunning{Euclid Collaboration: L. Pizzuti et al.}
   
   \maketitle

\section{\label{sc:Intro}Introduction}
Since the discovery of the late-time acceleration of the Universe (\citealt{Riess1998,Perlmutter1998}), the so-called $\Lambda$-cold-dark-matter (\lcdm) model has emerged as a concordance scenario able to describe a wide range of cosmological observations at different scales (e.g., \citealt{2020AA...641A...6P,2002MNRAS.330L..29E, 2004ApJ...606..702T,2006PhRvD..74l3507T,2005MNRAS.362..505C,Beutler2017}). %including the cosmic microwave background (CMB; e.g., \citealt{2020AA...641A...6P}), the galaxy power spectrum and baryon acoustic oscillations (BAO) in the large-scale structure (LSS; e.g., \citealt{2002MNRAS.330L..29E, 2004ApJ...606..702T,2006PhRvD..74l3507T,2005MNRAS.362..505C,Beutler2017}). 
In addition to baryonic matter and radiation, \lcdm assumes two dominant unseen components: cold dark matter (DM), which drives the formation and evolution of cosmic structures, and dark energy (DE), described in its simplest form by a cosmological constant $\Lambda$ in Einstein's equations of general relativity (GR). 
However, the fundamental nature of these components remains unknown. 

At the same time, increasingly precise observations have revealed tensions between early- and late-time probes \citep[e.g.,][]{2021CQGra..38o3001D,2022PhRvD.105b3520A,2023arXiv230611124L}. In addition, the new data release of the Dark Energy Spectroscopic Instrument \citep[DESI,][]{2016arXiv161100036D} shows a preference for dynamically evolving DE,
%with equation-of-state (EoS) parameters $w_0 > -1$ and $w_a < 0$,
when baryonic acoustic oscillations (BAO), supernovae (SNe), and cosmic microwave background (CMB) data are combined (\citealt{2025arXiv250314738D}). Whether these discrepancies arise from observational systematics or new physics remains debated, but they motivate the exploration of extensions to the standard paradigm.
A broad class of alternatives to \lcdm has therefore been developed. To address the unknown nature of $\Lambda$, DE models introduce new components with non-standard with equation-of-state \citep[EoS, e.g.,][]{Escamilla_2024}, while modified gravity scenarios alter GR on cosmological scales while recovering standard gravity locally (e.g., \citealt{Joyce_2016,Brax_2021}). Both possibilities generally involve additional degrees of freedom that affect structure formation at linear and non-linear scales (e.g., \citealt{PhysRevD.87.104015,Murgia_2016,Amendola_2019,2022MNRAS.513.1623C}), potentially leaving signatures detectable by ongoing and next-generation cosmological surveys.

Among Stage-IV facilities, \Euclid \citep{laureijs2011euclid,EuclidSkyOverview,2022AA...662A.112E} will provide an unprecedented wide-field dataset to probe the geometry and growth of LSS over nearly half the sky. Led by ESA with contributions from NASA, the mission employs a 1.2\,m telescope equipped with the visible imager VIS and the near-infrared spectrophotometer NISP, providing galaxy shapes, optical and NIR photometry \citep{2022A&A...657A..90E,2023A&A...671A.102E,2023A&A...671A.101E}, and slitless spectroscopy for redshift estimation \citep{Desprez-EP10,Ilbert-EP11} over the nominal $14\,000\,$deg$^2$ Wide Survey. These measurements will enable precise weak gravitational lensing (WL) and galaxy clustering analyses, improving constraints on DE, DM, and gravity \citep{Martinelli21b,2022A&A...660A..67N,2023A&A...671A.100E}. 
From the resulting photometric catalogue, galaxy clusters will be detected using dedicated algorithms such as \texttt{AMICO} \citep{Bellagamba2018,Adam-EP3} and \texttt{PZWav} \citep{Gonzalez2014,Thongkham24}. 

Cluster masses will be primarily calibrated through WL with the \texttt{COMB-CL} pipeline, which combines the high-precision VIS shape catalogue with cluster lensing measurements \citep{Cropper16,EuclidSkyVIS,EP-Sereno}. Although WL provides a powerful statistical mass calibration for large cluster samples, individual cluster masses are affected by intrinsic shape noise, contamination from unlensed galaxies, photometric-redshift uncertainties (\citealt{McClintock2019,EP-Lesci}), and secondary effects such as miscentring, triaxiality, and mergers (for instance, \citealt{2011ApJ...740...25B}, \citealt{Sommer2024} and \citealt{EP-Ingoglia}).

To fully exploit \Euclid sensitivity to non-standard signatures in cluster profiles, theoretical predictions from cosmological $N$-body simulations are crucial \citep[see][for a review]{Borgani2011}. These simulations follow the non-linear evolution of the matter density field under gravity, allowing structure formation to be studied in both \lcdm and alternative cosmologies (\citealt{Press:1973iz,1978IAUS...79..409Z,Potter2017,angulo2022review}). By tracing halo assembly and mass distributions across cosmic time, they provide predictions for departures from the standard model in the non-linear regime. In parallel, cluster detection algorithms \citep{Castignani2016,Gonzalez-Perez2014,Bellagamba2018} and inferred cluster properties \citep{Pacaud2007,Xu2018} are affected by projection effects \citep{Giocoli-EP30,EP-Ragagnin,Giocoli2025,Saxena2025,Cao2025}, requiring analyses of simulations that correctly account for interlopers and line-of-sight structures in matter density profiles. 

Previous studies on cosmological simulations in non-standard cosmologies have shown that halo profiles and related structural quantities carry non-trivial information on departures from $\Lambda$CDM. For instance, in interacting-dark-energy models, changes in the growth history and in the dark-sector dynamics alter halo density profiles and concentrations, leaving signatures in the internal structure of haloes beyond a simple change in the overall amplitude of structure formation \citep[e.g.][]{Baldi2012,Cui2012,Liu_2022,Zhao2025}. Massive neutrinos affect halo structure through a different physical mechanism: their free streaming suppresses the growth of structure, while their non-linear clustering around massive haloes adds a small, spatially extended contribution to cluster mass profiles. \cite{Villaescusa-Navarro2011} found that this component may induce a percent-level weak-lensing signature in massive cluster stacks, although its detection is strongly limited by baryonic feedback and projection systematics. More recently, for instance \cite{Hernandez_2024a} used high-resolution simulations to revisit neutrino halo profiles in more details and at lower neutrino masses, further showing that the neutrino component is extended, weakly clustered in low-mass systems, and potentially anisotropic around massive haloes.

Halo-scale signatures of modified gravity have also been extensively investigated with dedicated simulations. In screened scalar--tensor models such as $f(\mathcal{R})$ gravity \citep{Buchdahl:1970ynr}, the additional scalar degree of freedom generates a fifth force whose efficiency depends on halo mass, radius and environment. Simulations have shown that this force affects radial velocity profiles and the relation between dynamical and lensing masses \citep{Gronke2015,Gronke2016}, as well as halo concentrations and their connection to assembly histories \citep{Oleskiewicz2019}. The joint impact of $f(\mathcal{R})$ gravity and massive neutrinos has been first explored by \citet{Baldi2014}, who showed that neutrino free streaming can compensate the enhanced growth induced by modified gravity, and by \citet{Hagstotz:2019gsv}, who calibrated this degeneracy for cluster abundances. These results suggest that non-standard physics can leave profile-level signatures, and that neglecting them in \Euclid-like WL analyses may introduce cosmology-dependent systematic biases.

Within this context, this work aims to quantify, in a controlled DM-only setting, how selected classes of non-standard cosmologies modify the radial structure of massive haloes. %in view of the expected \Euclid cluster samples. 
We focus on three-dimensional density, cumulative-mass and velocity-dispersion profiles, together with projected mass and excess-surface-density profiles, as diagnostics of departures from the corresponding \lcdm predictions. To this end, we apply an extended version of the halo-profile reconstruction pipeline introduced in \cite{EP-Racz} to two simulation suites, \DUSTGRAINPATHFINDER \citep{Baldi2014} and \DEMNUni \citep{Carbone_2016} Taken together, these simulations sample three physically distinct mechanisms that can affect halo structure: the scale- and environment-dependent fifth force of $f(\mathcal{R})$ gravity, the suppression of structure growth and extended clustering induced by massive neutrinos, and the modified expansion and growth history associated with evolving dark energy. Although this model set is not exhaustive, it provides a physically motivated test-bed for assessing the size, radial dependence, and possible observational relevance of profile-level deviations from $\Lambda$CDM. We therefore adopt mass ranges and redshift cuts compatible with the \Euclid survey expectations \citep{Sartoris2016}. We further study the concentration--mass relation, which links halo mass to its internal density distribution and accretion history \citep{Maccio2007}, providing a crucial test of structure formation models in \lcdm and beyond \citep{Press:1973iz,Despali:2015yla,Liu_2022,Lopez-Cano2022}, and useful insights for modelling multi-wavelength observational data analyses \citep{Stanek2010,Farahi2019,Ragagnin2022}.

Note, however, that baryonic processes can modify the radial profiles of haloes in ways that may be partially degenerate with cosmological parameters such as the DE EoS \citep[e.g.][]{Copeland2018}.
While neglecting baryons is a strong simplification, DM-only simulations provide a necessary baseline for isolating the genuine impact of non-standard cosmologies on halo profile structure and for assessing whether such effects may be detectable by \Euclid-like surveys. This paper is structured as follows. In Sect.~\ref{sec:theo} we briefly describe the cosmological models investigated; in Sect.~\ref{sec:sim} we summarise the  main features of the simulations; and in Sect.~\ref{sec:method} we introduce the halo profiler. Section~\ref{sec:results} presents the main results, which are discussed in Sect.~\ref{sec:conclusions}. Throughout the paper, we consider proper distances, both three-dimensional and projected, unless explicitly stated.

%=================================================================================================================
\section{\label{sec:theo}Non-standard cosmological models} 
%=================================================================================================================
In this section, we present a brief overview of the alternatives to the $\Lambda$CDM scenario over which the two sets of simulations used in this paper are based. In particular, we will focus on dynamical dark energy $w$CDM cosmologies, $f(\mathcal{R})$ gravity, and neutrino cosmologies with different masses [both in $f(\mathcal{R})$ and in $\Lambda$CDM]. We refer to \cite{EP-Racz} for a more detailed description of the theoretical aspects.

%--------------------------------------------------------------------
\subsection{\texorpdfstring{$w$CDM}{wCDM} DE models}
%--------------------------------------------------------------------
The DE EoS may exhibit a redshift dependence rather than being strictly a constant.  
A widely used phenomenological parameterisation is the Chevallier--Polarski--Linder (CPL) form \citep{2001IJMPD..10..213C,PhysRevLett.90.091301},
\begin{equation}
  w_{\rm DE}(z) \;=\; w_0 + w_a \frac{z}{1+z}
  \;=\; w_0 + w_a\, (1 - a) \,,
  \label{eq:wzCPL}
\end{equation}
where \(w_0\) denotes the current EoS, and \(w_a\) quantifies its first-order evolution with the scale factor \(a\).  
This ansatz can be understood as a Taylor expansion of a smooth \(w_{\rm DE}(z)\) about \(z = 0\) (or \(a = 1\)), under the assumption of moderate evolution over cosmic time.  
Despite its simplicity, the CPL model can approximate the late-time expansion histories of a broad class of dynamical DE models with good accuracy \citep{2008GReGr..40..329L,2008PhRvD..78b3526L}.  
The special case \((w_0, w_a)=(-1,0)\) recovers the cosmological constant \(\Lambda\).

Recent observational constraints placed percent-level bounds on \((w_0, w_a)\), which are generally consistent with \lcdm predictions \citep[e.g.,][]{Brout_2022}.  
%In particular, the Pantheon+ compilation yields the following values
%\[
%w_0 = -0.978^{+0.024}_{-0.031}\,, \quad w_a = -0.65^{+0.28}_{-0.32}\,,
%\]
Interestingly, the combination of DESI DR1 BAO, CMB, and SNeIa has instead revealed a mild preference for evolving EoS, although the results remain compatible with \lcdm within uncertainties \citep{Lodha2025DESIphys,Adame2024DESI6}. 
These findings underscore that while a cosmological constant remains a viable null hypothesis, modest deviations in \(w(z)\) at low redshift remain observationally allowed. A modified expansion history and the corresponding change in the linear growth rate affect the time available for halo assembly, thereby changing mass-accretion histories, concentrations and the radial distribution of matter with respect to their \lcdm counterparts (Sect.~\ref{subsec:stacked_prof}).

%--------------------------------------------------------------------
\subsection{Massive relativistic neutrinos}
%--------------------------------------------------------------------
Massive neutrinos play a dual role in cosmology, influencing both the expansion history and the growth of cosmic structures. 
They are characterised primarily by two quantities: their total mass, $\Sigma\,m_{\nu}$ and the effective number of relativistic species, $N_{\mathrm{eff}}$. 
The latter parameter quantifies the contribution of all relativistic particles to the radiation energy density $\rho_{\mathrm{r}}$, via
\begin{equation}
\rho_{\mathrm{r}} = \left[1 + \frac{7}{8}\left(\frac{4}{11}\right)^{4/3} N_{\mathrm{eff}} \right]\rho_{\gamma} \, ,
\end{equation}
where $\rho_{\gamma}$ is the photon background energy density. 
Within the standard model of particle physics, three active neutrino species that thermalised in the early Universe lead to $N_{\mathrm{eff}} \simeq 3.045$ \citep{Cielo:2023bqp}. 
This value arises from the detailed treatment of neutrino decoupling, which includes non-instantaneous processes during $e^{+}e^{-}$ annihilation \citep{Mangano:2001iu}. 
A deviation from the fiducial $N_{\mathrm{eff}}$ would indicate the presence of additional relativistic relics or non-standard neutrino interactions. 
In what follows, we focus on standard active neutrinos only.

Neutrino oscillation experiments \citep{Maltoni:2004ei,Kajita:2016cak} have established that at least two mass eigenstates are non-zero, with measured mass-squared differences implying a minimum total mass of $\Sigma\, m_{\nu} \simeq 0.06\,{\rm eV}$ for the normal and $\Sigma\, m_{\nu} \simeq 0.10\,{\rm eV}$ for the inverted hierarchy. 
Cosmology constrains the sum of masses since gravity is sensitive to the total energy density of all species. 
The corresponding neutrino density parameter is
$
\Omega_{\nu} = (\Sigma \,m_{\nu})/(93.14\, h^{2}\, {\rm eV}) \,, 
$
and the total matter content reads $\Omega_{\mathrm{m}} = \Omega_{\mathrm{cdm}} + \Omega_{\mathrm{b}} + \Omega_{\nu}$, with $\Omega_{\mathrm{cdm}}$, $\Omega_{\mathrm{b}}$, and $\Omega_{\nu}$ the density parameter of cold dark matter (CDM), and baryons respectively.

Massive neutrinos decouple while still relativistic, just before Big Bang nucleosynthesis, and become non-relativistic at a redshift $1 + z_{\mathrm{nr}} \simeq 1890 \left( m_{\nu}/{\rm eV} \right)$,
when their temperature falls below their rest mass \citep{castorina_2015}. 
After this epoch, they behave as a non-relativistic matter component, contributing to the total matter density but with a large thermal velocity dispersion. 
For neutrino masses $m_{\nu} \lesssim 0.6\,{\rm eV}$, this transition occurs after recombination, thus affecting the late-time evolution of perturbations.

Free streaming of massive neutrinos suppresses the growth of structures below a characteristic comoving scale \citep{lesgourgues2006},
\begin{equation}
k_{\mathrm{fs}}(z) \simeq 0.82\, \frac{H(z)}{H_{0}\,(1+z)^{2}} 
\left( \frac{m_{\nu}}{{\rm eV}} \right) h~{\rm Mpc}^{-1} \,.
\end{equation}
For $k \gtrsim  k_{\mathrm{fs}}$, neutrinos do not cluster and do not contribute to gravitational potential wells \citep{takada2006}. 
On scales larger than the free-streaming horizon (i.e., $k \lesssim  k_{\mathrm{fs}}$) , neutrinos follow the CDM distribution, whereas on smaller scales their large velocities wash out density perturbations, leading to a scale-dependent suppression of the total matter power spectrum.

At linear order, this suppression can be approximated as
\begin{equation}
\frac{P_{\mathrm{mm}}(k; f_{\nu})}{P_{\mathrm{mm}}(k; f_{\nu}=0)} \simeq \xi\,f_{\nu} \, , \quad f_{\nu} = \frac{\Omega_{\nu}}{\Omega_{\mathrm{m}}} \,,
\end{equation}
where $\xi \in [1,8]$, with the CDM power spectrum suppressed roughly by a factor $1$--$6\,f_{\nu}$ \citep{castorina_2015}. At the non-linear level, \citet{Bird2011} showed that $\xi$ can reach a value of roughly 10 for the total matter power spectrum, increasing the sensitivity to the neutrino mass with respect to linear expectations. 
%Thus, even a small neutrino fraction $f_{\nu}\!\sim\!0.02$--$0.04$ produces a measurable reduction ($\sim$5--10\,\%) in the clustering amplitude on small scales.
As concerns the halo shapes, the scale-dependent suppression delays structure formation and can reduce the concentration and velocity dispersion of DM haloes at fixed mass. In addition, the non-linear clustering of the neutrino component around massive haloes produces a diffuse and extended contribution to the total matter distribution \citep[e.g.][]{Villaescusa-Navarro2011}.
%--------------------------------------------------------------------
\subsection{\texorpdfstring{$f(\mathcal{R})$}{f(\mathcal{R})} models of gravity}
%--------------------------------------------------------------------
In the context of modifications of GR, $f(\mathcal{R})$ gravity \citep{Buchdahl:1970ynr} is a sub-class of scalar-tensor theories viable at cosmological scales, equipped with a chameleon screening mechanism \citep{Khoury:2003aq}. It extends the Einstein--Hilbert action by introducing a general function $f(\mathcal{R})$ of the Ricci scalar $\mathcal{R}$,
\begin{align}
 S = \frac{c^4}{16\pi G_{\rm N}} \int{{\rm d}^4 x \,\sqrt{-g}\, \left[\,\mathcal{R}+f(\mathcal{R})\,\right]} \,, \label{eq:EHaction}
\end{align}
where $g$ is the determinant of the metric tensor $g_{\mu\nu}$, $G_\text{N}$ the gravitational constant and $c$ the speed of light. The quantity $f_R=\diff f/\diff R$ behaves as a new degree of freedom, the `scalaron', which is equivalent to a chameleon field in the Einstein frame, mediating a Yukawa-type fifth force that enhances gravity on scales larger than the screening radius (e.g., \citealt{2024Univ...10..443P}). Although this model has already been tightly constrained at both astrophysical and cosmological scales, it remains an interesting case study with rich phenomenology to understand how possible departures from GR affect cosmological observables.  

Here, we consider the Hu--Sawicki model \citep{Hu:2007nk} with $n=1$, implemented within the \DUSTGRAINPATHFINDER suite (see Sect.~\ref{ssec:DUSTGRAIN_sims}). In the limit of $f_R\ll1$, we have
\begin{equation}
f(\mathcal{R}) = - 6\, \ol\,\frac{H_0^2}{c^2} + |f_{\mathcal{R}0}|\,\frac{\bar{\mathcal{R}}_0^2}{\mathcal{R}}\,,\label{eq:fR}
\end{equation}
where $\bar R_0$ is the Ricci scalar evaluated at the background at the present time; $f_{\mathcal{R}0} = f_\mathcal{R}(z=0)$ is the corresponding background field value, $H_0$ is the Hubble constant, and $\ol$ is the energy-density parameter of the cosmological constant. The parameter $|f_{\mathcal{R}0}|$ characterises the magnitude of the deviation from \lcdm, with smaller values corresponding to weaker departures from GR until we recover \lcdm\ in the limit of $f_{\mathcal{R}0}\rightarrow0$. %For the small values of $|f_{\mathcal{R}0}|$ still allowed by observations, the background expansion history approximates that of \lcdm\ %and
%\begin{equation}
%\bar{\mathcal{R}}_0 = 3 \om \frac{H_0^2}{c^2} %\left(1+ 4 \frac{\ol}{\om} \right)\,,\label{eq:R}
%\end{equation}
%with the matter energy density parameter $\om=1-\ol$. 
Although the background expansion, for the small values of $|f_{\mathcal{R}0}|$ still allowed by observations, mimic that of the \lcdm model, it still differs at the level of cosmological perturbations, where the growth of structure is driven by a modification of gravity following the above adopted model of $f(\mathcal{R})$. The impact on halo profiles is therefore expected to be both dynamical and radial: in unscreened or partially screened regions the fifth force enhances infall and peculiar velocities, while the chameleon mechanism makes the deviations depend on halo mass, radius and environment.

The observational constraints on the model parameter $|f_{\mathcal{R}0}|$ vary from $|f_{\mathcal{R}0}| \lesssim 10^{-6}$ in the Solar System, $|f_{\mathcal{R}0}| \lesssim 10^{-8}$ from galaxy scales \citep{Burrage:2023eol} to $|f_{\mathcal{R}0}| \lesssim 10^{-6}$--$10^{-4}$ from various cosmological probes \citep[e.g., figure~28 in][for a summary]{Koyama:2015vza}. The parameter values of the simulations presented in this paper are similar to the current cosmological constraints.

%=================================================================================================================
\section{Overview of the cosmological simulations}
\label{sec:sim}
%=================================================================================================================
%\textcolor{red}{NOTE: Please read it and check that everything is consistent.}
Following the work of Euclid Collaboration: Carella et al. (in prep.), we focus on two cosmological $N$-body simulation sets implementing physics beyond the cosmological standard model; those sets are the \texttt{DEMNUni} and \texttt{DUSTGRAIN} simulations. In both suites, a flat $\Lambda$CDM run is performed along with its non-standard version(s).  We will compare profiles obtained in the non-standard frameworks with the corresponding $\Lambda$CDM case of the same simulation set to avoid problems related to different algorithms and numerical implementations. 

Note that the choice of the models analysed in this work is determined by the availability of homogeneous halo catalogues for our profile reconstruction. While not being an exhaustive sampling of the wide variety of interesting non-standard cosmologies, the selected models provide representative examples of the main physical mechanisms that can alter halo structure. The \DUSTGRAINPATHFINDER models isolate the impact of a chameleon-like fifth force in modified gravity frameworks, and its partial degeneracy with massive neutrino free streaming, while the \DEMNUni models probe the effects induced by dynamical dark energy and neutrino masses. %In this sense, the chosen scenarios are not necessarily extreme in terms of the parameter values allowed by current observations, but they are expected to produce among the largest profile-level deviations within the available suites, making them well suited to identify the radial and mass ranges where non-standard effects are most visible.
In the following we provide a brief summary of the simulation sets; for more details, see also Euclid Collaboration: Carella et al. (in prep.).

\subsection{\texttt{DUSTGRAIN} simulations}
\label{ssec:DUSTGRAIN_sims}

The \texttt{DUSTGRAIN} (Dark Universe Simulations to Test GRAvity In the presence of Neutrinos) project was designed to explore the degeneracy between modified gravity and massive neutrinos in the nonlinear regime of structure formation, first highlighted by \citet{Baldi2014}. 
It consists of two complementary suites of DM-only $N$-body simulations: \texttt{DUSTGRAIN-Pathfinder (PF)} \citep{Giocoli:2018gqh} and \texttt{DUSTGRAIN-fullscale}. 
Both employ the \texttt{MG-GADGET} code \citep{Puchwein:2013lza}, which self-consistently evolves $f(\mathcal{R})$ gravity, coupled with the implementation of massive neutrinos from \citet{Viel_2010} within the \texttt{P-GADGET3} framework. 
The performance of the modified gravity and neutrino modules has been validated and cross-checked against other numerical methods in \citet{Winther:2015wla} and \citet{Euclid:2022qde}. The recent \Euclid validation study of \citet{EP-Adamek} further compared \texttt{MG-GADGET} with several modern codes, finding percent-level agreement in the power-spectrum and in the relative halo mass function. This demonstrates that, although \texttt{MG-GADGET} is based on an older \texttt{GADGET} implementation, its modified gravity solver remains consistent with current state-of-the-art codes such as \texttt{MG-Arepo} \citep{2019NatAs...3..945A}.

As in Euclid Collaboration: Carella et al. (in prep.), in this work we focus on the \DUSTGRAINPATHFINDER series, designed to sample the joint $(f_{\mathcal{R}0},M_{\nu})$ parameter space and identify the most degenerate model combinations in LSS observables, such as the halo mass function \citep{Giocoli:2018gqh,Hagstotz:2018onp}. 
The suite comprises 13 cosmological runs with box size $L_{\mathrm{box}}=750\,h^{-1}$\,Mpc and $2\times768^{3}$ particles (for CDM and neutrino components). 
The reference $\Lambda$CDM cosmology (massless neutrinos) adopts $\Omega_{\rm m}=0.31345$, $\sigma_{8}=0.842$, $H_{0}=67.31$\,km\,s$^{-1}$\,Mpc$^{-1}$, and $n_{\rm s}=0.9658$, while the total matter density $\Omega_{\rm m}$ remains fixed when varying $m_{\nu}$. 
Each simulation provides 34 snapshots between $z_{\mathrm{in}}=99$ and $z=0$, in which the \rockstar halo finder \citep{Behroozi_2012} has been applied to identify the haloes and generate the catalogues used for our analysis. Here we consider three scenarios beyond \lcdm, namely a cosmology with $f_{\mathcal{R}0} = -1\times 10^{-5}$ [$f_5(\mathcal{R})$ hereafter], and two frameworks with the same $f_{\mathcal{R}0}$ value plus massive neutrinos with $m_\nu = 0.10$\,eV, and $0.15$ eV, respectively.

\begin{table*}
 \centering
 \caption{Cosmological parameter values of the \DUSTGRAINPATHFINDER simulation suite. 
 All models share the reference cosmology 
 $\{\Omega_{\rm b}, \Omega_{\rm m}, h, n_{\rm s}, A_{\rm s}\} = 
 \{0.0481, 0.31345, 0.6731, 0.9658, 2.199\times10^{-9}\}$. 
 Only parameters varying among models are reported below.}
 \label{tab:dpf_sets_table}
 \begin{tabular}{cccccccc}
 \hline
 \noalign{\vskip 3pt}
 {No.} & Type & $m_\nu$ [eV] & $\Omega_{\rm c}$ & $f_{\mathcal{R}0}$ & 
 $m_{\mathrm{cb}}^{\rm p}$ [$h^{-1}M_\odot$] & 
 $m_{\nu}^{\rm p}$ [$h^{-1}M_\odot$] & $\sigma_8$ \\
 \noalign{\vskip 3pt} 
 \hline
 \noalign{\vskip 3pt}
 {1} & $\Lambda$CDM & 0 & 0.26535 & 0 & $8.10\times10^{10}$ & 0 & 0.842 \\ 
 \noalign{\vskip 3pt}
 \hline
 % \textbf{2} & $\nu\Lambda$CDM & 0.15 & 0.26179 & 0 & $8.01\times10^{10}$ & $9.19\times10^{8}$ & 0.806 \\ \hline
 \noalign{\vskip 3pt}
 {2} & $f(\mathcal{R})$ & 0 & 0.26535 & $-1\times10^{-5}$ & $8.10\times 10^{10}$ & 0 & 0.903\\ 
 \noalign{\vskip 3pt}
 \hline
 \noalign{\vskip 3pt}
 {3} & \multirow{2}{*}{$\nu f(\mathcal{R})$} & 0.15 & 0.26179 & $-1\times10^{-5}$ & $8.01\times10^{10}$ & $9.19\times10^{8}$ & 0.864 \\
 {4} & & 0.10 & 0.26298 & $-1\times10^{-5}$ & $8.04\times10^{10}$ & $6.13\times10^{8}$ & 0.878\\ 
 \noalign{\vskip 3pt}
 \hline
 \end{tabular}
\end{table*}

\subsection{\label{ssec:DEMNUni_sims}\texttt{DEMNUni}}

The Dark Energy and Massive Neutrino Universe (\texttt{DEMNUni}) simulations \citep{Carbone_2016} were designed to explore the formation of the LSS in cosmologies including massive neutrinos and dynamical DE. They provide a coherent framework for nonlinear analyses of several probes, e.g., halo and galaxy clustering, WL, Sunyaev--Zeldovich effect, and cosmic void statistics \citep[e.g.,][]{castorina_2015,Vielzeuf_2022,Hernandez_2024a}.

The suite combines a large comoving volume with high-mass resolution, enabling simultaneous coverage of large- and small-scale perturbations. It comprises $15$ simulations, each with a volume of $8\,h^{-3}\,\mathrm{Gpc}^3$, softening length $\varepsilon=20\,h^{-1}\,\mathrm{kpc}$, and $2048^3$ CDM particles. In the runs with massive neutrinos, an additional $2048^3$ neutrino particles are included. The simulations start at redshift $z_{\rm in}=99$ with Zeldovich initial conditions, generated via a modified version of \texttt{N-GenIC} using Rayleigh-distributed amplitudes and random phases. The initial power spectrum is rescaled to $z_{\rm in}$ following the method of \citet{zennaro_2017}.

The runs were performed with a modified version of the TreePM-SPH code \texttt{P-Gadget3} \citep{Springel_2005}, adapted to evolve CDM and neutrino particles as separate collisionless species \citep{Viel_2010}. The reference cosmology corresponds to a flat $\Lambda$CDM model close to the \textit{Planck} 2013 best fit \citep{Planck_2013}; the CDM particle mass in the reference model is $m^{\rm p}_{\rm cb}=8.27\times10^{10}\,h^{-1}\,M_\odot$, decreasing proportionally with the neutrino mass to maintain a constant total matter density $\Omega_{\rm m}$. Note that massive neutrinos are implemented in a three-degenerate-mass approximation and treated as a single effective particle species. 

In Table \ref{tab:demnuni_sets_table} we list the non-standard cosmologies we employed for the current analysis. The full set of available simulations and the corresponding parameters can be found in Table 2 of Euclid Collaboration: Carella et al. (in prep.). As for the \DUSTGRAINPATHFINDER set, halo catalogues have been generated using the \texttt{Rockstar} halo finder. Here we consider only the available snapshot at $z=1.1$, using it as a high-redshift reference for \Euclid-oriented cluster analyses, rather than as a tomographic study of profile evolution, as further motivated in the next Section.

\begin{table*}
\centering
\caption{Cosmological parameter values of the \DEMNUni simulation suite. The reference cosmology for all models is $\{\Omega_{\rm b}, \Omega_{\rm m}, h, n_{\rm s}, A_{\rm s}\} = \{0.05, 0.32, 0.67, 0.96, 2.1265\times10^{-9}\}$. Only the parameters varied among the different scenarios are shown.}
% \begin{tabular}{|c|c|c|c|c|c|c|c|}
% \hline
% \textbf{No.} & Type & $m_\nu$ [eV] & $\Omega_{\rm c,0}$ & $(w_0,w_a)$ & $m_{\mathrm{cb}}^{p}$ [$h^{-1}M_\odot$] & $m_{\nu}^{p}$ [$h^{-1}M_\odot$] & $\sigma_8$ \\
% \hline
% \textbf{1} & $\Lambda$CDM & 0 & 0.2700 & $(-1,0)$ & $8.27\times10^{10}$ & 0 & 0.830\\
% \hline
% \textbf{2} & \multirow{2}{*}{$\nu\Lambda$CDM} & 0.16 & 0.2662 & \multirow{2}{*}{$(-1,0)$} & $8.17\times10^{10}$ & $9.97\times10^{8}$ & 0.793\\
% \textbf{3} & & 0.32 & 0.2623 & & $8.07\times10^{10}$ & $1.99\times10^{9}$ & 0.752\\
% \hline
% \textbf{4}–\textbf{7} & $w_0w_a$CDM & 0 & 0.2700 & varied & $8.27\times10^{10}$ & 0 & 0.78–0.86\\
% \textbf{8}–\textbf{15} & $\nu w_0w_a$CDM & 0.16–0.32 & 0.2662–0.2623 & varied & $8.17–8.07\times10^{10}$ & $9.97\times10^{8}–1.99\times10^{9}$ & 0.70–0.82\\
% \hline
% \end{tabular}
\begin{tabular}{rllccccc}
    \hline
    \noalign{\vskip 3pt}
    {No.} & Type & $m_\nu$ [eV] & $\Omega_{\rm c}$ & $(w_0,w_a)$ & $m_{\mathrm{cb}}^{\rm p}$ [$h^{-1}M_\odot$] & $m_{\nu}^{\rm p}$ [$h^{-1}M_\odot$] & $\sigma_8$ \\
    \noalign{\vskip 3pt}
    \hline
    \noalign{\vskip 3pt}
    {1} & \lcdm & 0 & 0.2700 & $(-1, 0)$ & $8.27 \times 10^{10}$ & 0 & 0.830\\
    \noalign{\vskip 3pt}
    \hline
    \noalign{\vskip 3pt}
    {2} & $\nu$\lcdm & 0.16 & 0.2662 & $(-1, 0)$ & $8.17 \times 10^{10}$ & $9.97 \times 10^8$ & 0.793\\
    \noalign{\vskip 3pt}
    \hline
    \noalign{\vskip 3pt}
    \multirow{4}{*}{{3}–{6}}
        & \multirow{4}{*}{$w_0w_a$CDM} 
            & \multirow{4}{*}{0}
                & \multirow{4}{*}{0.2700}
                    & $(-0.9, -0.3)$ 
                        & \multirow{4}{*}{$8.27 \times 10^{10}$} 
                            & \multirow{4}{*}{0} 
                                & 0.828\\
    
        &   &   &   & $(-0.9, +0.3)$ &   &   & 0.777\\
        &   &   &   & $(-1.1, -0.3)$ &   &   & 0.861\\
        &   &   &   & $(-1.1, +0.3)$ &   &   & 0.831\\
    \noalign{\vskip 3pt}
    \hline
\end{tabular}
\label{tab:demnuni_sets_table}
\end{table*}

\subsection{Methodology}
\label{sec:method}
%To extract the density and velocity dispersion profiles from haloes in each simulation's snapshot, we implement a module which is part of the pipeline described in \cite{EP-Racz}. %We first run the \rockstar halo finder, which is based on a friends-of-friends (FoF) algorithm that combines information from both positions and velocities of particles, and we consider the custom \texttt{BGC2} binary data format files, which are among the possible outputs of \rockstar. These files contain information about each halo, such as velocities, positions, and total mass, and the related particle data. Additional useful information is also stored, such as redshift, box length, and particle mass. This will serve as input for the routine that computes the halo profiles.
%For the illustrative purposes of this work, and given the limited availability of the \DEMNUni snapshots, in each simulation we consider as our reference analysis a single snapshot at redshift $z=1.1$, which is close to the peak of the number counts distribution within the range expected from the Euclid Wide and Deep Surveys \citep{Sartoris2016}. 
%Moreover, different \Euclid forecast and validation studies indicate that the WL signal-to-noise ratio progressively degrades with increasing lens redshift and becomes significantly lower at $z \gtrsim 1$ \citep{Giocoli-EP30,EP-Ingoglia}.
%In this context, the redshift $z=1.1$ can be regarded as a conservative, near-limiting case for WL profile constraints with \Euclid.
The halo density profiles analysed in this work are generated with the parallel \texttt{Python3} algorithm \texttt{GetHaloDensityProfiles}. This routine is part of the pipeline first presented in \cite{EP-Racz} and it will be made publicly available upon completion of the full pipeline. Starting from the halo catalogs, the code constructs binned profiles only for haloes resolved with a minimum number of particles $N_{\rm min}$, to ensure that all derived quantities are adequately sampled. We adopt a threshold $N_{\rm min}=1000$ particles and compute profiles using 50 logarithmically spaced radial bins over the range $[0.001,\,5]\,r_{500{\rm c}}$ for all haloes in the sample.\footnote{
We define $r_{\Delta{\rm c}}$ as the radius enclosing an average density $\Delta_{\rm c}\,\rho_{\rm cr}(z)$, where $\rho_{\rm cr}(z)$ is the critical density of the Universe at redshift $z$, and $M_{\Delta{\rm c}}$ as the corresponding enclosed mass.}

\texttt{GetHaloDensityProfiles.py} outputs three-dimensional cumulative mass, mass density, number density, and cumulative number density profiles, together with velocity and velocity-dispersion profiles for both the Cartesian and spherical components of the velocity field.
In addition, the code produces 2-dimensional projected profiles by integrating the particle distribution along each Cartesian axis within a cylindrical volume of half-length $5\,r_{500{\rm c}}$, thus providing three independent lines of sight for each halo. This choice represents a trade-off between capturing the full one-halo contribution to the projected profiles and minimising contamination from LSS unrelated to the halo. For NFW-like density profiles, the contribution from radii larger than a few times $r_{500{\rm c}}$ rapidly becomes negligible, thus further extending the analysis primarily adds noise rather than a physically meaningful signal.
In this work, we focus our main analysis to a single snapshot at redshift $z=1.1$. While this choice was initially guided by practical constraints regarding the availability of the halo catalogs across all cosmological models -- in particular for the \DEMNUni set -- this specific redshift is in fact highly relevant for \Euclid WL science. As shown by \cite{Sartoris2016}, $z=1.1$ probes a key epoch for \Euclid observations, marking a critical transition regime where the WL signal-to-noise ratio begins to degrade significantly \citep{Giocoli-EP30,EP-Ingoglia}. Analysing synthetic data at $z=1.1$ therefore provides a demanding and conservative stress test of the cosmological model impact on halo profiles under challenging high-redshift observational conditions, meaning that our results can be considered a conservative lower bound on our ability to distinguish between these varying cosmological frameworks. For \DUSTGRAINPATHFINDER, where snapshots at lower $z$ are available, we further consider haloes at $z = 0.5, 0.3$ to assess how the non-standard signatures change at later times.

%All profiles are stored in Hierarchical Data Format version~5 (HDF5). A subsequent post-processing step is then applied to extract the physical quantities relevant for the analysis presented in this work, namely, the halo concentration parameters, the velocity anisotropy profiles, and the excess surface density profiles. 
%
We restrict the analysis to haloes with mass larger than $10^{13}\,M_{\odot}$. Although the \Euclid survey selection function implies that only clusters with masses
$\gtrsim 10^{14}\,M_{\odot}$ are expected to be robustly detected \citep{Sartoris2016}, we adopt a lower mass threshold as a compromise between observational relevance and the need to maximise the statistical power of the halo sample. We note that any effects induced by the finite simulation resolution can be neglected when interpreting the results obtained from this work: as shown by~\cite{Power_2003}, the halo internal mass distribution can only be robustly traced down to a given resolution-dependent convergence radius -- corresponding to the radius at which the average collisional relaxation time approximately equals 0.6 times the age of the Universe -- below which the profiles may be significantly affected by resolution. In the case of \DEMNUni and \DUSTGRAINPATHFINDER, this unresolved region corresponds to radii $r \lesssim 0.1 \,r_\mathrm{500c}$ for haloes with $M_\mathrm{200c} \simeq 10^{13}\, M_\odot$, and becomes smaller at larger masses. Therefore, we only conduct our profile analysis for radii $r \gtrsim 0.1\, r_\mathrm{500c}$ in order to minimize the impact from the simulation resolution.

\section{Results}
\label{sec:results}
In this section, we elaborate on the analyses we have performed on the reconstructed profiles and the main results obtained for the \DEMNUni and \DUSTGRAINPATHFINDER sets. In particular, we report stacked halo mass, density, velocity dispersion spherical components ($\sigma_{ r}$, $\sigma_\theta$ and $\sigma_\phi$) and velocity anisotropy profiles $\beta(r) = 1 - (\sigma_\theta^2 +\sigma_\phi^2)/(2\,\sigma^2_{ r}) $ profiles; then we explore the concentration--mass relation. We stress again that our study is based on DM-only simulations, %Therefore, the profile differences discussed below should not be considered as full observational forecasts for \Euclid clusters, but as 
with the aim of providing a controlled baseline for the differential impact of non-standard cosmological physics on halo structure. This caveat is important because baryonic processes -- such as gas cooling, star formation, and feedback from SNe and active galactic nuclei -- can modify matter profiles at a level comparable to, or even larger than, the deviations induced by changes in cosmology, especially in the inner regions of haloes. At the same time, the DM-only approach is well suited to the main purpose of this work. By comparing each non-standard model to its corresponding \lcdm reference simulation within the same suite and numerical setup, we minimise purely numerical differences and characterise the differential response of modification of gravity, the presence of massive neutrinos, and a dynamical dark energy, on the radial matter distribution of haloes. In modified gravity scenarios, for instance, the fifth force acts on all non-relativistic components and can alter both DM and baryonic dynamics, while baryons may in turn modify the efficiency of screening through their contribution to the local gravitational potential. Hydrodynamical simulations will therefore be required for fully realistic predictions, but the DM-only comparison provides the necessary reference for identifying where non-standard effects are expected to be largest before adding baryonic and observational systematics.

A related caveat concerns the interpretation of the differences between the $f(\mathcal{R})$ and $\Lambda$CDM models in the \DUSTGRAINPATHFINDER suite. Despite starting from the same initial conditions, the $f_5(\mathcal{R})$ run has a larger late-time clustering amplitude than its $\Lambda$CDM counterpart, with $\sigma_8=0.903$ instead of $\sigma_8=0.842$ (Table~\ref{tab:dpf_sets_table}). Part of the difference in the halo profiles can therefore arise from the enhanced growth amplitude and the corresponding shift of the halo mass function, rather than from the fifth force alone. Nevertheless, the chameleon mechanism introduces a characteristic mass-, environment-, and radius-dependent response: screened halo cores and massive systems are expected to react differently from low-density outskirts and lower-mass haloes. This scale-dependent behaviour is not equivalent to a simple global rescaling of the fluctuation amplitude, and it motivates the profile-based comparison carried out below. 

More generally, standard cosmological parameters affect halo profiles through their impact on formation times, accretion histories, and concentrations \citep[e.g.][]{2018ApJS..239...35D}. Since these parameters are currently constrained at the percent level in the baseline $\Lambda$CDM model \citep{2020AA...641A...6P}, standard concentration--mass models suggest that the induced profile variations should be of order percent to few percent. Such variations are likely smaller than the largest $f(\mathcal{R})$ residuals found here, but may be comparable to the weaker signatures induced by massive neutrinos and evolving dark energy. A quantitative propagation of these uncertainties to the profile residuals would require matched simulations, or an emulator, in which cosmological and non-standard parameters are varied independently. Since the present work does not perform a cosmological likelihood analysis, we leave this marginalisation to future work.

\subsection{\label{subsec:stacked_prof}Stacked halo profiles} 
To obtain a statistically robust estimate of the impact of non-standard cosmologies on the halo profiles, we stacked the quantities of interest (3D and 2D mass and density, velocity dispersion, and velocity anisotropy profiles), considering three mass bins, in order to study how the effects of alternative models shape the profiles at different mass scales.
To ensure having roughly the same number of haloes in each mass bin for every model, we first compute the virial mass $M_{\rm 200c} = 200 H^2(z) / (2G_{\rm N})\, r_{200c}^3$, being the mass of a sphere of radius $r_{200c}$. %enclosing an average density 200 times the critical density of the Universe at redshift $z$. 
For each cosmology, we then choose the bin edges to be the $33\,\%$ and $66\,\%$ quantiles of each $M_{\rm 200c}$ distribution, in order to have, in each bin, the same abundance of haloes for every cosmology. These intervals roughly correspond to the \lcdm masses $M_{\rm 200c}/M_\odot \in \left[ 10^{13},\ 6.8 \times 10^{13} \right)$, $M_{\rm 200c}/M_\odot \in \left[6.8 \times 10^{13},\ 10^{14} \right)$ and $M_{\rm 200c}/M_\odot \in \left[ 10^{14},\ 3 \times 10^{15} \right]$. %for each model, and result in each mass bin being populated by similar amounts of haloes regardless of cosmology.
% The projected mass and density excess profiles (see below) are reported in Fig.~\ref{fig:profiles_2Dx} for both suites, while the stacked 3D halo profiles are shown in Figs.~\ref{fig:DEMNUni_stacked_profiles_3D} for \DEMNUni and \ref{fig:DUSTGRAIN_stacked_profiles_3D} for \DUSTGRAINPATHFINDER.

The stacking is performed by computing both mean and median halo profiles for each mass bin, and along the $x, y, z$ directions for the projected profiles. We also show the relative differences between each non-standard model and \lcdm in the bottom panel of each plot. We further compute the surface density excess profiles defined as $\Delta \Sigma (R) = \Sigma (<R) - \Sigma(R)$ along each axis, where $\Sigma (<R)
 = M_{\rm proj}(<R)/ (\pi R^2)$ is the average projected mass density within $R$. This quantity is connected to the tangential shear generated by a weak gravitational lensing signal of a massive halo -- one of the main observables that will be measured by \Euclid (e.g., \citealt{EP-Ingoglia}). Indeed, the reduced tangential shear $g_\text{t}$ of a source at redshift $z_\text{s}$, produced by a lens at $z_{\rm d}$ is proportional to the projected mass and density profiles as 
\begin{equation} \label{eq:tan_shear}
    g_\text{t} = \frac{\gamma_\text{t}}{1-\kappa}\,,\,\,\, \text{where}\,\,\,\, \kappa(R) = \frac{\Sigma(R)}{\Sigma_{\rm crit} }\,, \quad 
   \gamma_{\rm t}(R) = \frac{\Delta \Sigma (R)}{\Sigma_{\rm crit}}\,.
\end{equation}
%where 
%\begin{equation}
%   \kappa(R) = \frac{\Sigma(R)}{\Sigma_{\rm crit} }\,, \quad 
%   \gamma_{\rm t}(R) = \frac{\Delta \Sigma (R)}{\Sigma_{\rm crit}}\,.
%\end{equation}
Above, the critical surface density is defined as
\begin{equation}
    \Sigma_{\rm crit} \equiv \frac{c^2}{4\pi G_{\rm N}}\frac{D_{\rm s}}{D_{\rm l}\, D_{\rm ls}}\,,
\end{equation}
where $D_{\rm l}(z_{\rm d})$, $D_{\rm s}(z_{\rm s})$, and $D_{\rm ls}(z_{\rm d},z_{\rm s})$ are the lens, source, and lens-source angular diameter distances.

It should be noticed that in $f(\mathcal{R})$ gravity, the lensing potential is equivalent to the standard Newtonian potential, due to the conformal structure of the theory (see, e.g., \citealt{2022Univ....8..157P} and references therein); however, the shape of the surface density profile encodes the information about the new degree of freedom, which alters the assembling history of haloes. In more general models where the relativistic sector is also modified (e.g., more general Horndeski: \citealt{Kofinas_2017}, \citealt{Kobayashi_2019}; beyond Horndeski frameworks: \citealt{Gleyzes_2015}, \citealt{Dima_2018}), the WL signal is no more a direct probe of matter distribution, and the contribution of the fifth force may produce enhanced effects on the reconstructed WL profiles.

\begin{figure*}[ht]
    \centering
    \centerline{
    \includegraphics[width=0.42\hsize]{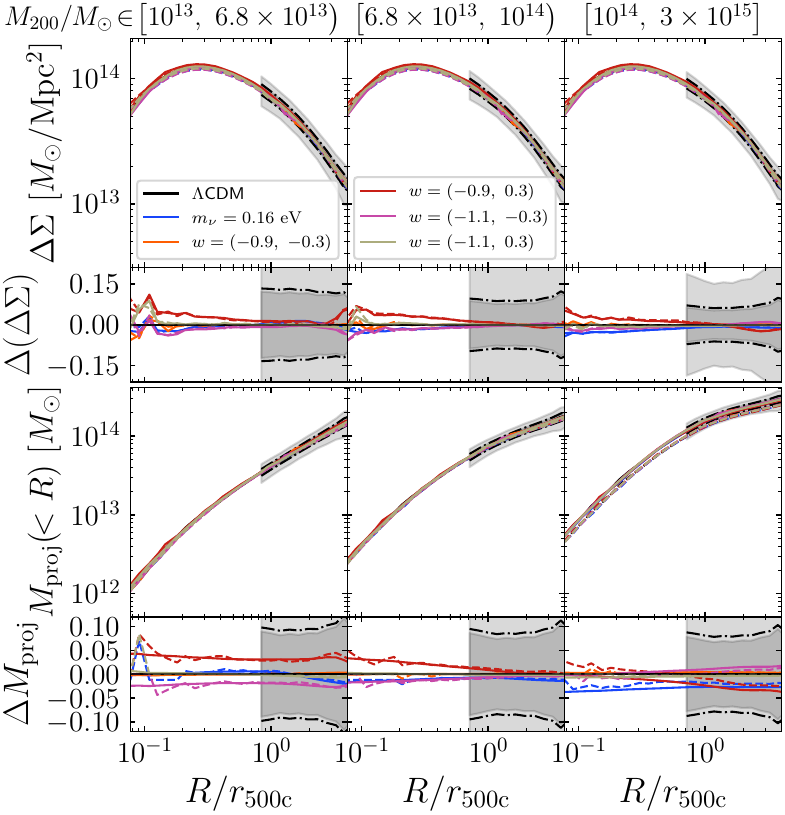}
    \includegraphics[width=0.42\hsize]{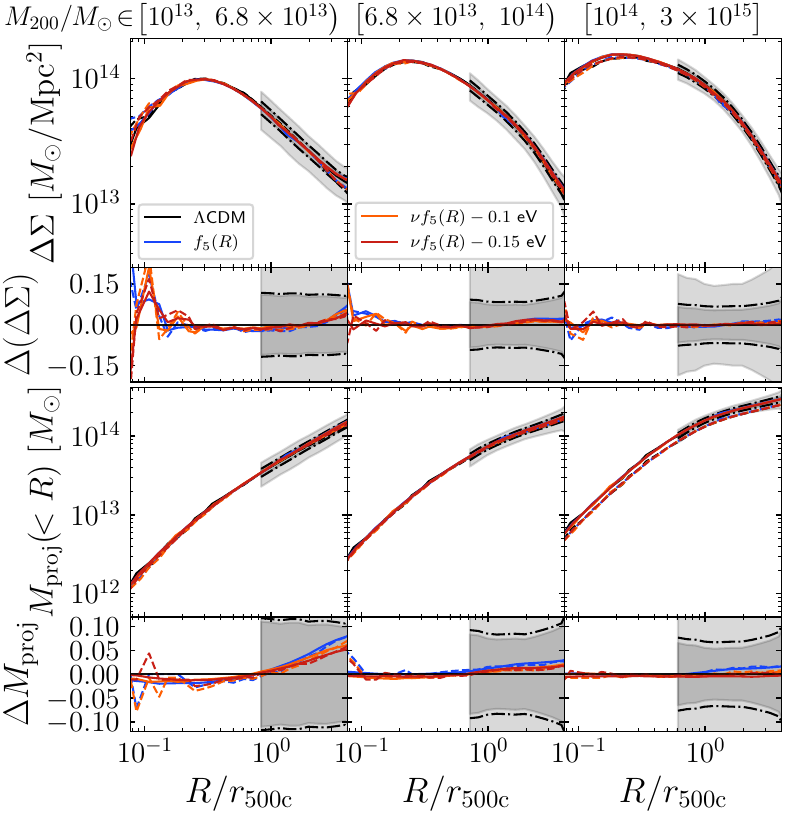}}
    \caption{Stacked projected profiles along the $x$-axis for each mass interval, at $z = 1.1$. \textit{Top panels}: \DEMNUni and \DUSTGRAINPATHFINDER projected surface density excess profiles. \textit{Bottom panels}: projected mass profiles. Each solid line is a mean profile, and the dashed line of the same colour is the corresponding median profile. Each colour corresponds to a different model, and the relative differences are computed with respect to the \lcdm model [for a quantity $X$, we define $\Delta X \equiv (X - X_\mathrm{\Lambda CDM})/X_\mathrm{\Lambda CDM}$]. The grey and dark grey regions in the surface density excess plots are the \Euclid uncertainty bands predicted when stacking $N = 10^3, 10^4$ haloes, respectively, considering a $f_{\rm sys} = 0.03$. We also show the \Euclid uncertainties for $f_{\rm sys} = 0.05$ and $N = 10^4$ as the black dot-dashed line.} %Note the switching of the $(w_0, w_a) = (-0.9, 0.3)$ and $(w_0, w_a) = (-1.1, -0.3)$ profiles when passing from low to high masses.}
    \label{fig:profiles_2Dx}
\end{figure*}

Figure \ref{fig:profiles_2Dx} shows the projected mass and surface density excess along the $x$-axis (2Dx, hereafter) for \DEMNUni and \DUSTGRAINPATHFINDER. We do not show the $y$ and $z$ projections as they do not deviate particularly from the 2Dx profiles, implying that any eventual triaxiality and orientation effects in the individual halo profiles are washed out by the stacking procedure.
Overall, the relative differences between mean and median projected profiles of different frameworks hardly exceed about $10\,\%$ except possibly at radii lower than about $0.1\,r_{500c}$ (which are, as stated above, excluded from the analysis). %However, since lower radius bins tend to contain very few particles ($N \lesssim 100$), they are dominated by numerical noise and should not be treated as concrete evidence of any non-standard effects. 
In detail, no significant differences are detectable in the $(w_0, w_a) = (-0.9, -0.3)$ cosmology, while $(w_0, w_a) = (-1.1, 0.3)$ deviates only by less than $1\,\%$. Similarly, the mass profiles in the $\nu \Lambda$CDM cosmology remain un-altered for low to intermediate masses, displaying the growth suppression of neutrinos only in the larger mass bin. Instead, our slowest-accelerating model $(w_0, w_a) = (-0.9, 0.3)$ results in amplified mass profiles with respect to \lcdm, with deviations of up to $5\%$, before becoming lower than \lcdm at greater masses. On the contrary, the fastest-accelerating model $(w_0, w_a) = (-1.1, -0.3)$ produces projected mass profiles that are consistently lower than \lcdm, except in the high-mass interval where the profile becomes larger. A similar situation is found in the density excess profiles. This behaviour can be understood in terms of the different growth histories associated with the various cosmological models, despite identical initial conditions. In scenarios where the linear growth factor is suppressed, such as models with massive neutrinos or dynamical dark energy (e.g., \citealt{Verza_2023}), structure formation is delayed, resulting in a halo mass function that is shifted towards lower masses with respect to $\Lambda$CDM. As a consequence, haloes populating a given mass bin at a fixed $z$ in these cosmologies typically correspond to less massive progenitors in the $\Lambda$CDM case (see, e.g., Fig.~10 of \citealt{castorina_2015}), which are also expected to be more concentrated. 
Thus, the projected mass and density profiles of low-mass haloes in these cosmologies are suppressed at all considered radii. Conversely, the effect of their larger concentration becomes more relevant at higher masses, as \lcdm haloes are less concentrated and thus contain less particles in each radial bin, leading to an enhancement of the profiles in non-standard models. An analogous reasoning is valid for the slower-accelerating cosmologies, where the linear growth factor is boosted instead of being suppressed, implying a halo mass function favoring higher masses at a given redshift.
Meanwhile, the suppression of growth in massive-neutrino cosmologies becomes apparent only in the largest mass bins. 
%We note that these trends are further affected by the finite mass resolution of the simulations, which limits the sampling of low-mass haloes and, in the case of massive neutrinos, leads to an additional suppression of the halo mass function at small masses. Probing these effects more robustly would require simulations with higher resolution and access to lower-mass haloes.

With regards to the \DUSTGRAINPATHFINDER simulations, the impact of modified gravity and massive neutrinos is most pronounced in the outer regions of dark matter haloes, where the fifth force of the $f_5(\mathcal{R})$ model is only weakly screened. As shown in the bottom left panels of Fig.~\ref{fig:profiles_2Dx}, the projected mass profiles exhibit a mild enhancement at large radii, particularly for low-mass haloes, consistently with an increased late-time accretion driven by the fifth force in low-density environments. The inclusion of massive neutrinos partially counteracts this effect, since neutrinos preferentially cluster in the halo outskirts and enhance the screening efficiency, bringing the profiles closer to the \lcdm expectation \citep[e.g.,][]{Baldi2014,Hagstotz:2018onp}. 

The surface density excess profiles further support this picture, being largely indistinguishable from \lcdm at intermediate and high halo masses, and displaying deviations of at most $\sim 5\,\%$ only in the outskirts of low-mass systems. This behaviour can be understood in terms of the different growth histories induced by modified gravity: haloes selected at fixed mass and redshift in $f(\mathcal{R})$ cosmologies typically correspond to less massive progenitors in \lcdm, having experienced a more efficient and prolonged accretion phase in their outer regions. As a result, the mass is redistributed towards larger radii, mildly reducing the inferred concentration and enhancing the projected mass profile in the halo outskirts. In this framework, the presence of massive neutrinos suppresses the fifth-force-driven growth, effectively compensating the modified gravity effects and yielding profiles that remain close to the \lcdm case.

Since WL serves as the primary \Euclid observable, we also compare the stacked profiles with the predicted \Euclid uncertainties on the surface density excess profiles at $z_{\rm d} \sim 1.1$. 
Analogously to the analysis of \citet{EP-Ingoglia} and \citet{EP-Sereno}, we assume that, for a single halo, the average observed surface density excess in the $i$-th radial bin is computed by weighted average of the reduced tangential shear (Eq.~\ref{eq:tan_shear}) generated by background sources in the bin. 
%and therefore we define the corresponding density excess profile uncertainty as \citet{EP-Ragagnin}: 
%
%\begin{equation} \label{eq:dsigma_error}
%	\delta \langle\Delta \Sigma_{g_{\rm t}}\rangle_i = \Sigma_{\rm crit} \frac{\sigma_\varepsilon}{\sqrt{\pi n_{\rm g} \left( R_{i + 1}^2 - R_i^2 \right)}}\,.
%\end{equation}
%Above, $n_{\rm g}$ is the number density of background sources, $\sigma_\varepsilon$ is the dispersion of intrinsic source shapes, and $R_i$ is the inner edge of the $i$-th projected radial bin. Similarly to \cite{EP-Ingoglia} we restrict the error analysis to the radial interval $R \in \left[ 0.3,\ 5 \right]$ Mpc, corresponding to the range in which WL is best detected. For our purposes, at $z = 1.1$ we compute $\Sigma_{\rm crit} \simeq 1.77 \times 10^{15} \, \mathrm{M_\odot}\,{\rm Mpc}^{-2}$, while $n_{\rm g} \simeq 14\ \text{arcmin}^{-2}$. We further assume that $\sigma_\varepsilon \simeq 0.3$ \citep{EuclidSkyOverview,EP-Ragagnin}.

We first define the corresponding shape-noise contribution to the density excess profile uncertainty as \citet{EP-Ragagnin}:
\begin{equation}
\label{eq:dsigma_shape_error}
\delta \langle\Delta \Sigma_{g_{\rm t}}\rangle_{i,{\rm shape}}
=
\Sigma_{\rm crit,eff}(z_{\rm d})\,
\frac{\sigma_\varepsilon}
{\sqrt{n_{\rm g}(z_{\rm d}) \,A_i}}\,.
\end{equation}
Here $n_{\rm g}(z_{\rm d})$ is the effective number density of background sources, $\sigma_\varepsilon$ is the dispersion of intrinsic source shapes, and $A_i
=\pi
\left(
\theta_{i+1}^2
-
\theta_i^2
\right) ,$ is the angular area of the $i$-th annulus, with
$
\theta_i = R_i/D_{\rm A}(z_{\rm d})\,.
$
Similarly to \cite{EP-Ingoglia} we restrict the error analysis to the radial interval $R \in \left[ 0.3,\ 5 \right]$ Mpc, corresponding to the range in which WL is best detected. The effective critical density, $\Sigma_{\rm crit,eff}(z_{\rm d})$, is obtained by averaging over the adopted source redshift distribution, as described below.

For the source redshift distribution, we adopt the \Euclid-like Smail parametrisation commonly used in WL forecast studies,
\begin{equation}
p(z_s)
=
\mathcal{N}
\left(\frac{z_s}{z_0}\right)^2
\exp\left[
-\left(\frac{z_s}{z_0}\right)^{3/2}
\right] \,.
\end{equation}
The normalisation $\mathcal{N}$ is fixed by requiring the integral of $p(z_s)$ over redshift to be unity. Following the \Euclid forecast convention, we set $z_0=z_{\rm m}/\sqrt{2}$, with median source redshift $z_{\rm m}=0.9$ \citep[e.g.][]{laureijs2011euclid,Deshpande-EP28}. For a lens at redshift $z_{\rm d} = 1.1$, we compute
\begin{equation}\label{eq:effdens}
\left\langle \Sigma_{\rm crit}^{-1}(z_{\rm d}) \right\rangle
=
\int_{z_{\rm d}}^{\infty}
{\rm d}z_{\rm s}\,
p(z_{\rm s}|z_{\rm s}>z_{\rm d})\,
\Sigma_{\rm crit}^{-1}(z_{\rm d},z_{\rm s}) \,,
\end{equation}
where $p(z_s|z_s>z_{\rm d})$ is the source redshift distribution renormalised over the background population. We then define $\Sigma_{\rm crit,eff}(z_{\rm d})$ as the inverse of Eq.~\eqref{eq:effdens}. The shape-noise term in Eq.~\eqref{eq:dsigma_shape_error} represents the optimistic statistical uncertainty associated with source ellipticities. To account in a simplified, effective way for additional uncertainties that are not explicitly modelled in our dark-matter-only analysis, we introduce two further contributions. The first is a stochastic intrinsic-profile scatter,
\begin{equation}
\delta \langle\Delta \Sigma_{g_{\rm t}}\rangle_{i,{\rm int}}
=
f_{\rm int}
\left|
\Delta\Sigma_{\Lambda{\rm CDM}}(R_i)
\right| \,,
\end{equation}
which parametrises cluster-to-cluster variations, triaxiality and projection effects. Since this term represents random scatter among different haloes, it decreases as $N_{\rm cl}^{-1/2}$ in stacked profiles. The second contribution is a coherent systematic floor,
\begin{equation}
\delta \langle\Delta \Sigma_{g_{\rm t}}\rangle_{i,{\rm sys}}
=
f_{\rm sys}
\left|
\Delta\Sigma_{\Lambda{\rm CDM}}(R_i)
\right|\, ,
\end{equation}
which accounts for residual shear calibration, photometric-redshift calibration, source-selection effects, miscentring, and profile-modelling uncertainties. Differently from the former, this term represents a coherent residual uncertainty, thus it is not reduced by stacking. The total uncertainty is therefore

\begingroup
\small
\begin{equation} \label{eq:euclid_dsigma_error}
\left[
\delta \langle\Delta \Sigma_{g_{\rm t}}\rangle_{i,{\rm tot}}
\right]^2 =
\frac{
\left[
\delta \langle\Delta \Sigma_{g_{\rm t}}\rangle_{i,{\rm shape}}
\right]^2
+
\left[
\delta \langle\Delta \Sigma_{g_{\rm t}}\rangle_{i,{\rm int}}
\right]^2
}
{N_{\rm cl}} +
\left[
\delta \langle\Delta \Sigma_{g_{\rm t}}\rangle_{i,{\rm sys}}
\right]^2 \,.
\end{equation}
\endgroup

As fiducial values for the magnitude coefficients, we adopt $f_{\rm int}=0.20$ and $f_{\rm sys}=0.03$. We also consider $f_{\rm sys}=0.05$ as a more conservative stress test. These values are motivated by recent \Euclid weak-lensing simulation studies, which show that triaxiality and orientation can induce tens-of-percent variations in individual cluster lensing masses, while residual modelling assumptions typically lead to average weak-lensing mass biases of a few percent, reaching values between $5\%$ and $10\%$ in less constrained cases \citep{Giocoli-EP30}. %In this sense, $f_{\rm int}$ represents the stochastic component of cluster-to-cluster profile scatter, whereas $f_{\rm sys}$ represents a coherent residual uncertainty that cannot be beaten down by increasing the number of stacked clusters.
For each non-standard model, we quantify the radial residual with respect to the corresponding $\Lambda$CDM prediction as
\begin{equation}
\Delta[{\Delta\Sigma}(R_i)]
=
\left[\Delta\Sigma_{\rm model}(R_i)
-
\Delta\Sigma_{\Lambda{\rm CDM}}(R_i)\right]/ \Delta\Sigma_{\Lambda{\rm CDM}}(R_i)\,.
\end{equation}
We then compute the cumulative significance of the residual as
\begin{comment}
\begin{equation} \label{eq:SN_ratio}
\left(\frac{S}{N}\right)^2
=
\sum_i
\frac{
\left[
\Delta\Sigma_{\rm model}(R_i)
-
\Delta\Sigma_{\Lambda{\rm CDM}}(R_i)
\right]^2
}
{
\left[
\delta \langle\Delta \Sigma_{g_{\rm t}}\rangle_{i,{\rm tot}}(N_{\rm cl})
\right]^2
}\,,
\end{equation}
\end{comment}
\begin{equation} \label{eq:SN_ratio}
\left(\frac{S}{N}\right)^2
=
\sum_i \left[
\Delta\Sigma_{\rm model}(R_i)
-
\Delta\Sigma_{\Lambda{\rm CDM}}(R_i)
\right]^2/
\left[
\delta \langle\Delta \Sigma_{g_{\rm t}}\rangle_{i,{\rm tot}}
\right]^2
\,,
\end{equation}
which is used to estimate the number of stacked clusters required to reach a detection with target significance, $N_{\rm req}$. We report the values in Table~\ref{tab:N_req_table}, for a significance at $3\sigma$,  whenever they are finite. If the coherent systematic floor prevents the target significance from being reached even in the large-$N_{\rm cl}$ limit, we classify the corresponding profile residual as systematics-limited.

\begin{figure*}%[ht]
    \centering
    \centerline{
    \includegraphics[width=0.45\hsize]{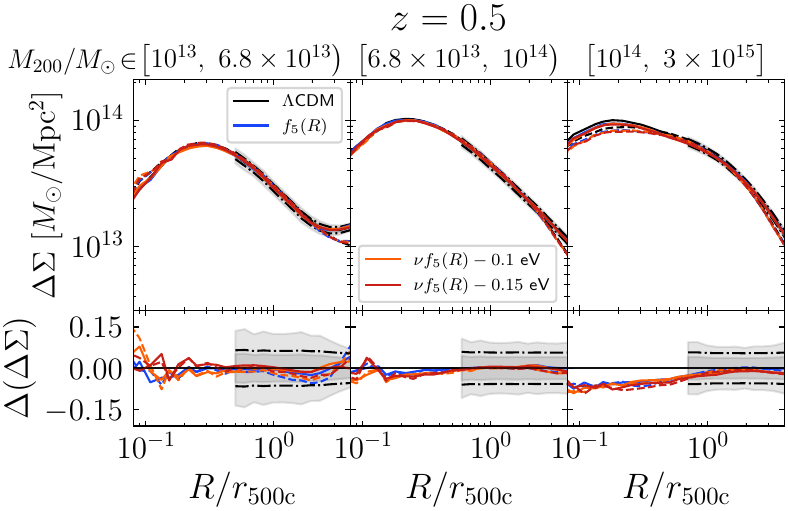}
    \includegraphics[width=0.45\hsize]{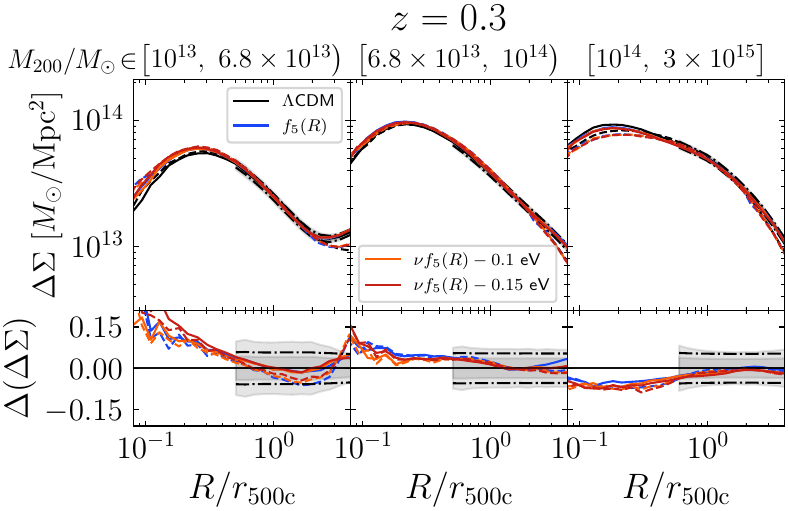}}
    \caption{Excess surface density profiles for two snapshots of the \DUSTGRAINPATHFINDER suite, at $z=0.5$ (left) and $z = 0.3$ (right). The colour code is the same as in Fig.~\ref{fig:profiles_2Dx}.}
    %with grey and dark grey bands representing the expected \Euclid uncertainties for $10^3$ and $10^4$ (stacked) clusters, respectively, considering $f_{\rm sys} = 0.03$. We also show the \Euclid uncertainties for $f_{\rm sys} = 0.05$ and $N = 10^4$ as the black dot-dashed line.}
    \label{fig:Excess_2Dx0502}
\end{figure*}

The grey shaded regions in Fig.~\ref{fig:profiles_2Dx} show the predicted density excess uncertainties expected from \Euclid WL observations, computed using Eq.~\eqref{eq:euclid_dsigma_error}, in the case of $N_{\rm cl} \in \{10^3,\ 10^4\}$ stacked clusters. We assume $n_{\rm g}(1.1) = 14\ {\rm arcmin^{-2}}$, $f_{\rm int} = 0.2$, and $f_{\rm sys} = 0.03$; we also show the conservative case with $f_{\rm sys} = 0.05$ for $N_{\rm cl} = 10^4$.
We find that, for both simulation suites, the effects of non-standard cosmologies on the WL profiles at $z = 1.1$ remain below the expected \Euclid precision in both cases considered. Indeed, stacking with $10^4$ haloes only allows reduction of the uncertainties at the $\sim 8\%$ level, whereas deviations from \lcdm barely exceed $\sim 5\%$. We predict that the required number of haloes needed to detect such effects with a $3\sigma$ significance through lensing stacking analyses is $\sim 10^5$, which at present day remains very optimistic (see Appendix~\ref{app:detectability}). Our analysis thus suggests that \Euclid-like WL profiles may be unsuitable to discriminate cosmological scenarios for clusters at $z \gtrsim 1$. However, the WL signal is expected to increase significantly for lenses at lower redshift; therefore, it is interesting to explore whether the \Euclid error bars for surface density excess profiles can decrease enough at $z < 1$ to allow for detection of non-standard effects with a lower number of clusters. In order to provide an indication in this regard, we consider two additional snapshots in the \DUSTGRAINPATHFINDER suite, one at $z = 0.5$ and another at $z=0.3$.
In Fig.~\ref{fig:Excess_2Dx0502} we show the excess surface densities obtained from the analysis of these snapshots (top $z=0.5$, bottom $z= 0.3$), compared with the expected \Euclid uncertainties. Following figure~4 of \cite{EuclidSkyOverview} and Table 1 of \cite{Giocoli-EP30}, we consider $n_{\rm g}(0.5) = 26\ {\rm arcmin^{-2}}$ and $n_{\rm g}(0.3) = 28\ {\rm arcmin^{-2}}$.

At these lower redshifts, the deviations from \lcdm become more pronounced and display a clear dependence on halo mass. In the most massive haloes, the modified gravity profiles tend to be suppressed in the inner regions and enhanced in the outskirts relative to \lcdm, indicating a lower effective concentration. This behaviour can be interpreted as a consequence of chameleon screening: while the dense cores of massive haloes remain efficiently screened, their outer regions are still partially sensitive to the fifth force, which boosts late-time accretion and redistributes matter towards larger radii.

Conversely, lower-mass haloes show the opposite tendency, with enhanced central profiles and weaker or reversed deviations in the outskirts. In this mass regime screening is less effective, allowing the fifth force to operate over a larger fraction of the halo volume and promote a more efficient collapse of the inner regions, leading to profiles that appear more centrally concentrated than in \lcdm.

The inclusion of massive neutrinos produces only a mild modification of these trends. While neutrinos suppress the overall growth of structure, their effect does not fully compensate the nonlinear and radius-dependent response induced by modified gravity within haloes, particularly at low redshift where the fifth-force effects have had more time to accumulate.
Interestingly, the enhancement of the non-standard effects (up to more than $10\%$) and the slight decrease of the \Euclid uncertainties at low redshifts demonstrate that a few thousand clusters in the low- and intermediate-mass bins may now be sufficient to detect signatures of new physics in the stacked WL signal (see Appendix.~\ref{app:detectability}).

From a complementary perspective, even when not directly detectable, profile modifications induced by departures from the \lcdm scenario can still act as non-negligible, cosmology-dependent systematics in WL analyses. If the WL profiles generated in a non-standard cosmology are interpreted with rigid \lcdm-based templates, coherent radius-dependent deviations in $\Delta\Sigma(R)$ can be partially absorbed into the inferred shape of the profile, which may produce a biased calibration of the halo mass. We do not propagate this effect to cosmological parameters here, since this would require a full likelihood analysis including dedicated forward modelling of observational and astrophysical systematics, %cluster selection, baryons, projection effects, miscentring, and photometric-redshift uncertainties, 
which is however beyond the purposes of this paper. Instead, our results identify the mass, redshift, and radial regimes where non-standard profile modifications are large enough to require explicit modelling in future \Euclid cluster analyses.
%if the true underlying cosmology departs from standard $\Lambda$CDM, about $10\,\%$ unmodelled modifications in the halo profiles induced by non-standard physics may translate into a non-negligible systematic bias in cosmological constraints inferred from \Euclid WL analyses. 
%Note, however, that this comparison should be considered cautiously as an order-of-magnitude estimate. Indeed, for the sake of simplicity here we are considering only statistical errors; a realistic detectability assessment will require dedicated forward modelling of observational and astrophysical systematics, which is however beyond the purposes of this paper.

Regarding the 3D profiles (see Figs.~\ref{fig:DEMNUni_stacked_profiles_3D} and~\ref{fig:DUSTGRAIN_stacked_profiles_3D}), the \DEMNUni cosmologies principally alter the mass and velocity dispersion, while leaving the density only slightly affected. The largest deviation of the mass profiles is found at lower masses, approximately at the $5\%$ level at all radii. Instead, high-mass haloes present much more significant differences in their velocity dispersions at large radii, exceeding $5\%$ in the case of the radial dispersion. 
On the other hand, the impact of the \DUSTGRAINPATHFINDER non-standard models is particularly evident in the density and velocity dispersion profiles of haloes in the lowest mass bin, exceeding $10\%$ and reaching up to $20\%$ differences in the velocity dispersions. The suppression of the $f_5(\mathcal{R})$ model by neutrinos is still evident in a similar way to the 2D profiles. 
%{\bf 
%It is interesting to note that, for medium- to high-mass haloes, the behaviour of the green and red models switch, as a result of the effects on halo concentrations due to their different growth histories. We will address detectability in Appendix~\ref{app:detectability}.
%}

The velocity anisotropy profiles $\beta(r)$ remain mostly unaltered among the various models tested.  For low-mass haloes, a weak trend of lower $\beta(r)$ profiles is found in $f(\mathcal{R})$ gravity, corresponding to more isotropic orbits. This confirms our findings from the density profiles: haloes that, at a given redshift and mass, have experienced a longer accretion are likely to be more virialised compared to their $\Lambda$CDM counterpart, and this translates into $|\beta|<1$. The slight ``isotropisation'' of the orbits in $f(\mathcal{R})$ is also found in the low-$z$ snapshots, but the effects remain always smaller than $10\%$ (see Fig.~\ref{fig:DUSTGRAIN_beta_low_z}). 

Notably, in every cosmological model of both simulation suites the radial profiles $\beta(r)$ tend to drop significantly around $2\, r_{500}$, after an initial increase with radius. Several studies on orbits in massive systems, both from simulations (e.g., \citealt{2021A&A...652A..90A}) and observations (e.g., \citealt{2019A&A...631A.131M,2025A&A...699A..88P}) confirm that $\beta(r)$ grows on average from the centre up to the virial radius, indicating a preference of more radial orbits $\beta > 0$ in the outskirt where galaxies (particles) are still infalling into the potential well of the halo. On the other hand, the sharp decrease of $\beta(r)$ at larger radii, particularly evident in the intermediate mass bins of both sets, points towards the dominance of tangential orbits.  
This behaviour has been discussed in \cite{Abdullah2025}, found to be driven by the simultaneous presence of galaxies on infalling trajectories and galaxies that have already undergone pericentric passage and are moving outward~\citep[thus will be connected to the splashback radius,][]{DiemerKravstov2014}. The overlap of these orbital populations leads to a redistribution of angular momentum, which manifests as an increase in the tangential velocity dispersion.

%================================================================
\subsection{\label{subsec:cm}Concentration--mass relation} 
%================================================================
The concentration of a dark matter halo is one of its most fundamental properties, representing an estimate of the compactness of the halo, linked to its formation and assembling history (e.g., \citealt{2002ApJ...568...52W}). We define it as  $c_{200} \equiv r_{\rm 200c}/r_{-2}\,$, where $r_{-2}$ is the radius at which the logarithmic slope of the halo profile equals $-2$. Low-mass CDM haloes are expected to be denser and more concentrated on average than high-mass ones, giving rise to a decreasing trend of concentration with total mass. Thus, we aim to assess whether our non-standard cosmologies significantly affect this trend.

\begin{figure}
    \centering
    \includegraphics[width=0.8\columnwidth]{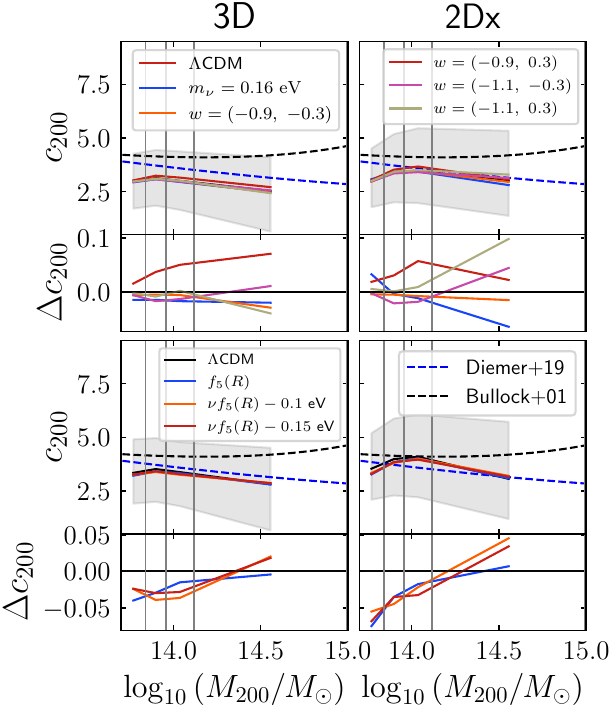}
    \caption{Median concentration--mass relations for each \DEMNUni (top) and \DUSTGRAINPATHFINDER (bottom) model and relative differences with \lcdm%(where $\Delta c_{200} \equiv (c_{200} - c_\mathrm{200, \Lambda CDM})/c_\mathrm{200, \Lambda CDM}$).
    . The \lcdm scatter is also shown as the shaded region, and the grey vertical lines represent the mass bin edges. We also show the theoretical predictions by~\cite{Bullock_2001} (black) and~\cite{Diemer_2019} (blue) as the dashed curves. Overall, every non-standard cosmology is only different from \lcdm at around the $10\%$ level: not significant enough to be detected through \Euclid WL data.}
    \label{fig:cM_rel_diff}
\end{figure}

 To this end, we compute the concentration--mass relation using the values of $c_{200}$ and $M_{\rm 200c}$ obtained by fitting the 3D and 2D cumulative mass distribution of every halo with Navarro--Frenk--White \citep[NFW,][]{Navarro_1997} model. The NFW density profile is a widely-adopted model found to adequately describe the radial matter distribution of haloes in cosmological simulations across several scales, as well as of observed galaxy clusters. The density of a NFW model is defined as
\begin{equation} \label{eq:NFW_profile}
    \rho_{\rm NFW}(r) = \frac{\rho_0}{x (1 + x)^2}\,,
\end{equation}
where $x \equiv r/r_{-2}$, and $\rho_0(r_{\rm 200c},r_{-2})$ is a characteristic density of the halo which depends on $r_{-2}$ and $r_{\rm 200c}$. We restrict the fitting range to $r/r_{500c}\in[0.2,1.5]$. The lower limit is chosen to lie safely above the estimated convergence radius discussed in Sect.~\ref{sec:method}, while also avoiding the innermost region where numerical resolution and, in real clusters, baryonic physics would be most relevant. The upper limit avoids giving excessive weight to the outer regions, where departures from equilibrium, splashback-related features, and neighbouring structures can make a single NFW description less appropriate. Note that we do not impose a relaxation selection in our sample. This choice is motivated by the fact that our goal is not to calibrate a universal concentration--mass relation for relaxed haloes, but to compare the same full halo populations used in the stacked profile analysis. Splitting the samples according to relaxation indicators, such as substructure mass fraction or virial ratio, would require homogeneous dynamical diagnostics for all catalogues and would introduce an additional, potentially cosmology-dependent selection.
The full concentration--mass relations are shown in
Figs.~\ref{fig:DEMNUni_cM_relation} and~\ref{fig:DUSTGRAIN_cM_relation} for \DEMNUni and \DUSTGRAINPATHFINDER, respectively. Due to the large intrinsic scatter of the samples, we compute a robust estimate of the $c_{200}$ trend by dividing all haloes into four $\logten{M_{\rm 200c}}$ bins, then computing the median concentration in each, along with the corresponding $68\%$ scatter. The bins are chosen to be populated by the same number of haloes, to minimise statistical fluctuations. 
%We noticed that haloes with unusually low concentration in both 3D and 2D cases (i.e. $c_{200}$ < 1) account for less than $15\,\%$ of the sample: we checked in a few cases that such unphysical values of $c_{200}$ are associated to very perturbed haloes in merging phase. 
The evolution of the median concentration with mass for all the \DEMNUni (top) and \DUSTGRAINPATHFINDER (bottom) cosmologies considered is shown in the top panels of each plot in Fig.~\ref{fig:cM_rel_diff}. The bottom panels in Fig.~\ref{fig:cM_rel_diff} further present the relative differences between the median concentrations with respect to \lcdm in each mass bin. Overall, all $f(\mathcal{R})$ deviations from \lcdm are within approximately the $5\%$ level in both the 3D and 2Dx cases.
Moreover, the effect of modified gravity is greater when considering lower-mass bins, exhibiting less concentrated haloes than \lcdm. Indeed, this is consistent with our analysis of the $f(\mathcal{R})$ stacked profiles in Sect.~\ref{subsec:stacked_prof}: the altered accretion history of haloes redistributes the halo mass towards the outskirts, and the effect is lowered by the presence of neutrinos. Then, the suppression of growth of massive neutrino cosmologies leaves less massive, but more concentrated haloes than \lcdm in the high-mass bins, boosting the median concentration. Notably, the median concentration--mass relation shows greater deviations from \lcdm (about $10\%$) when considering the low-redshift \DUSTGRAINPATHFINDER snapshots at $z = 0.5$ and $z = 0.3$, shown in Fig.~\ref{fig:DUSTGRAIN_cM_low_z}. Furthermore, in these snapshots we  observe an increase in the concentration of low-mass haloes with respect to the $z = 1.1$ case for all modified gravity models, while massive ones exhibit a lower $c_{200}$: this behaviour is indeed a further validation of what found in Sect.~\ref{subsec:stacked_prof}. 

Similarly, all \DEMNUni cosmologies only present less than about $10\%$ differences, with the most deviation found at large masses in the $(w_0, w_a) = (-0.9, 0.3)$ model for the 3D case, and in the $(w_0, w_a) = (-1.1, 0.3)$ in the 2Dx case. Moreover, the massive neutrinos cosmology presents a greater deviation in the large-mass bin. 
Overall, the relation between concentration and mass does not seem to exhibit noticeable enough differences (less than about $5\%$) among the various cosmologies analyzed, especially when considering its large intrinsic scatter. Indeed, following the analysis conducted in \cite{Giocoli2025}, these non-standard effects prove to be too subtle for detection using the concentration--mass relation derived from \Euclid WL observations. Thus, we do not find any strong evidence to allow distinction between non-standard models and \lcdm.

\section{Conclusions}
\label{sec:conclusions}

In this work, we have presented a systematic analysis of the internal structure of DM haloes in selected non-standard cosmological scenarios, focusing on the shape of their 3-dimensional and projected mass, density, and velocity dispersion profiles. Using an extended version of the pipeline introduced in \cite{EP-Racz}, we reconstructed halo profiles from two cosmological $N$-body simulation suites, \DEMNUni\ and \DUSTGRAINPATHFINDER, specifically designed to probe the effects of dynamical dark energy, massive neutrinos, and modified gravity beyond the \lcdm paradigm.

Our results show that halo profiles may constitute a sensitive probe of non-standard cosmological physics, particularly in the outer regions of haloes. In the \DUSTGRAINPATHFINDER\ simulations, $f(\mathcal{R})$ gravity produces systematic modifications of the mass and density profiles, which become more pronounced toward large radii. At $z\gtrsim 1$ these deviations remain relatively small, while at lower redshift they develop a clear dependence on halo mass. In the most massive systems the profiles tend to be suppressed in the inner regions and enhanced in the outskirts, indicating a lower effective concentration, whereas lower-mass haloes can display the opposite trend as a consequence of the reduced efficiency of chameleon screening. These deviations reach about the $10\%$ level beyond approximately $2\,r_{\rm 500c}$ in some mass bins. The inclusion of massive neutrinos generally reduces the amplitude of the modified gravity effects through the suppression of structure growth, although it does not completely erase the radius- and mass-dependent response induced by the fifth force in the nonlinear regime \citep{Baldi2014}. Similar trends are observed in the velocity dispersion profiles, while velocity anisotropies remain comparatively less affected, indicating a partial compensation between radial and tangential motions up to about $2\,r_{\rm 500c}$.\\
In the \DEMNUni\ suite, deviations from $\Lambda$CDM induced by dynamical dark energy and massive neutrinos are generally milder, at the few-percent level, but remain distinguishable in stacked mass and velocity dispersion profiles, particularly for low-mass haloes and at large radii. These effects can be consistently interpreted in terms of different growth histories: models with suppressed linear growth produce haloes that, at fixed mass and redshift, correspond to less massive progenitors in $\Lambda$CDM, leading to enhanced concentrations and outer mass profiles. Conversely, scenarios with accelerated structure formation display reduced concentrations and redistributed mass toward the outskirts. 

Note that, while in this work we have focused on the overall structure of the halo profiles, a particularly sensitive probe of modifications in the outskirts is the splashback radius $r_{\rm sp}$ \citep{More_2015}. Since $r_{\rm sp}$ is associated with the sharp steepening of the density profile near the first apocentre, even modest variations in the outer profile can induce non-negligible shifts in its location. However, the interpretation of this feature is not unique: the inferred splashback radius depends on the tracer population used to define the profile  \citep[e.g.,][]{Lebeau2024}. A robust characterization of $r_{\rm sp}$ in non-standard cosmologies would thus require dedicated modelling of the steepening region for the relevant observational tracers, and is left to future work.

We further investigated the concentration--mass relation by comparing the median $c_{200}$ profiles in log-spaced bins of mass for the various models; we found deviations of order $10\%$ which correctly reflect the effects on the mass and density profiles. However, such deviations are too small to be detected with high significance, given the scatter expected by \Euclid-like surveys.

From an observational perspective, our analysis provides a simulation-based framework for the interpretation of stacked WL measurements of galaxy clusters with \Euclid. %While lensing observables are not directly affected by the fifth force in $f(\mathcal{R})$ gravity, the imprint of modified growth histories is encoded in the shape of the projected mass and density profiles. 
Deviations from \lcdm depend on halo mass and redshift, with typical amplitudes of a few percent in the regimes explored here. In this context, our results indicate that stacking techniques will be essential to overcome observational noise. Adopting a simple toy-model for systematics we showed that detecting such effects with \Euclid\ data will require a large number, $\sim10^4$--$10^5$, of WL profiles at the largest redshift analysed. Although this may appear as a major limitation, we note that $z\sim 1$ represents a threshold close to a ``worst-case scenario'', above which the WL signal rapidly degrades. For $z\lesssim 1$, the number density of background sources increases and the critical surface density decreases, leading to a significantly improved signal-to-noise ratio. Moreover, the effects of non-standard cosmologies are expected to strengthen toward lower redshift; the analysis of the \DUSTGRAINPATHFINDER suite at $z =0.3,0.5$ showed a potential increase in the chances of detecting new physics with $\gtrsim 10^{3}$ stacked haloes, which is highly feasible given the capabilities of the \Euclid survey.
Beyond their detectability, these effects are directly relevant for \Euclid cluster cosmology.  Interpreting weak-lensing measurements with \lcdm-based profile templates when the true cosmology is non-standard may therefore introduce a cosmology-dependent modelling error in the inferred masses and concentrations. The size of this effect cannot be inferred from a single fractional number, since it depends on the radial shape of the residuals, halo mass, redshift, sample size, and the adopted systematic floor. Our \Euclid-oriented benchmark provides a first quantitative estimate of where such residuals are statistically relevant, while their propagation to cosmological constraints will require a dedicated forward model.

Future refinements will explicitly account for the diverse factors contributing to $f_{\rm sys}$ and $f_{\rm int}$. 
%We stress again that the comparison with \Euclid uncertainties presented here is intended as an idealised benchmark based on shape noise and a simplified modelling of systematics. A more robust assessment of the detectability of modified gravity signatures with \Euclid\ will require dedicated forward modelling explicitly accounting for several contributions.
%It should not be interpreted as a full forecast for the detectability of non-standard cosmological signatures in \Euclid stacked WL profiles, as it does not explicitly include several observational systematics that affect WL measurements. 
Along with the already mentioned projection effects \citep{Giocoli-EP30,EP-Ragagnin}, halo triaxiality and orientation bias \citep{Meneghetti2013,Meneghetti2014}, correlations with dynamical state \citep{Ludlow2013,Ludlow2014}, and optically inferred richness \citep{Ragagnin2022}, or miscentering \citep{Sommer2024} between the assumed halo centre and the true potential minimum, further modify stacked $\Delta\Sigma$ profiles at a level comparable to the residuals investigated here, and therefore need to be modelled jointly with the cosmology-dependent modifications discussed in this work. %A more robust assessment of the detectability of modified gravity signatures with \Euclid\ will therefore require dedicated forward modelling including these effects. 
%Finally, at high $z$, kinematic mass profile measurements will complement WL analyses, even though the low signal-to-noise ratio and the impact of systematics remain important limitations (Euclid Collaboration: Pizzuti et al., in prep.).
%Furthermore, it is worth noting that the trends discussed here are affected by the finite mass resolution of the simulations, particularly at the low-mass end and in massive-neutrino cosmologies, where the halo mass function is suppressed. Extending this analysis to higher-resolution simulations and a broader mass range will be crucial to fully assess the detectability of these effects. 

Finally, future works will extend the present analysis to hydrodynamical simulations, to assess the signature of non-$\Lambda$CDM physics on the observable baryonic content. Although the effects  of baryons are expected to be stronger in galaxy cluster' cores -- which we excised in this study study -- baryonic feedback can influence the matter distribution up to the virial radius and beyond \citep[e.g.,][]{Angelinelli2022,Angelinelli2023}. The combination of the DM-only baseline with the more complex analysis of full-physics simulation will provide robust predictions on alternative models, essential to isolate non-standard effects in the upcoming \Euclid observations and provide new, stringent constraints on the nature of the fundamental constituents of the Universe.
%it is important to point out again that we have focused on DM-only simulations. However, it is well known that baryonic physics significantly contributes in shaping the profiles of cosmic structures. Although the effects are expected to be stronger in galaxy cluster' cores -- which we excised in this study -- baryonic feedback can influence the matter distribution up to the virial radius and beyond \citep[e.g.,][]{Angelinelli2022,Angelinelli2023}. Future works will extend the present analysis to hydrodynamical simulations, to assess the signature of non-$\Lambda$CDM physics on the observable baryonic content. This will provide robust predictions on alternative models, essential to isolate non-standard effects in the upcoming \Euclid observations and provide new, stringent constraints on the nature of the fundamental constituents of the Universe.

\begin{acknowledgements}
LP and AR acknowledge the usage of INAF-OATs IT framework \citep{Taffoni2020,Bertocco2020}.
LP acknowledges support from the Italian Ministry for Research and University (MUR) under Grant ‘Progetto Dipartimenti di Eccellenza 2023–2027’ (BiCoQ).
AR acknowledges EuroHPC Joint Undertaking for awarding the project ID EHPC-REG-2024R01-029  access to Leonardo at CINECA, Italy.
AR acknowledges ISCRA for awarding this project access to the LEONARDO supercomputer, owned by the EuroHPC Joint Undertaking, hosted by CINECA, Italy (HP10BUFI59). 
The \texttt{DEMNUni} simulations were carried out in the framework of ``The Dark Energy and Massive-Neutrino Universe" project, using the Tier-0 IBM BG/Q Fermi machine and the Tier-0 Intel OmniPath Cluster Marconi-A1 of the Centro Interuniversitario del Nord-Est per il Calcolo Elettronico (CINECA). CC acknowledges a generous CPU and storage allocation by the Italian Super-Computing Resource Allocation (ISCRA) as well as from the coordination of the ``Accordo Quadro MoU per lo svolgimento di attività congiunta di ricerca Nuove frontiere in Astrofisica: HPC e Data Exploration di nuova generazione'', together with storage from INFN-CNAF and INAF-IA2. CTM is supported by an appointment to the NASA Postdoctoral Program at the NASA Goddard Space Flight Center, administered by Oak Ridge Associated Universities under contract with NASA.
 %\AckQone \AckQtwo %\AckERO
  \AckDatalabs
  \AckEC \AckCosmoHub \,See the reference papers \cite{Carretero2017,Tallada2020}.
\end{acknowledgements}

%
% Here comes the reference list, generated via bibtex from
% the bibfile AandA.bib
%

\bibliography{Euclid, DR1, Myref}

%
% You can have different .bib files here, e.g.
% \bibliography{Euclid,myown}, but please use the bibentries
% in Euclid.bib for all EC publications to ensure that these
% references are up-to-date and in correct format.
%

%
% Now you can add appendices.
% In this example, the appendices are in one column mode.
% If that is not requires, comment out \onecolumn
%

\begin{appendix}
  \onecolumn %If you don't want single column for the Appendix, please
\section{Three-dimensional profiles}
\begin{figure*}[h]
    \centering
    \centerline{
    \includegraphics[width=0.41\hsize]{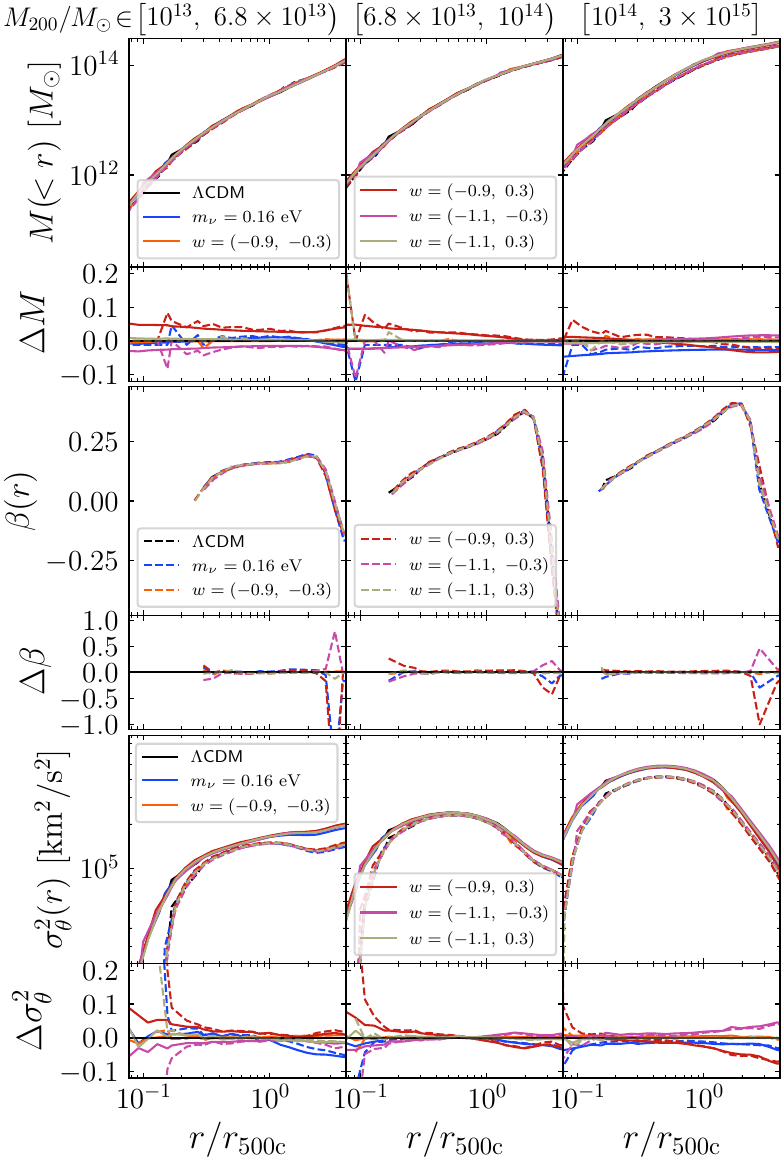}
    \includegraphics[width=0.41\hsize]{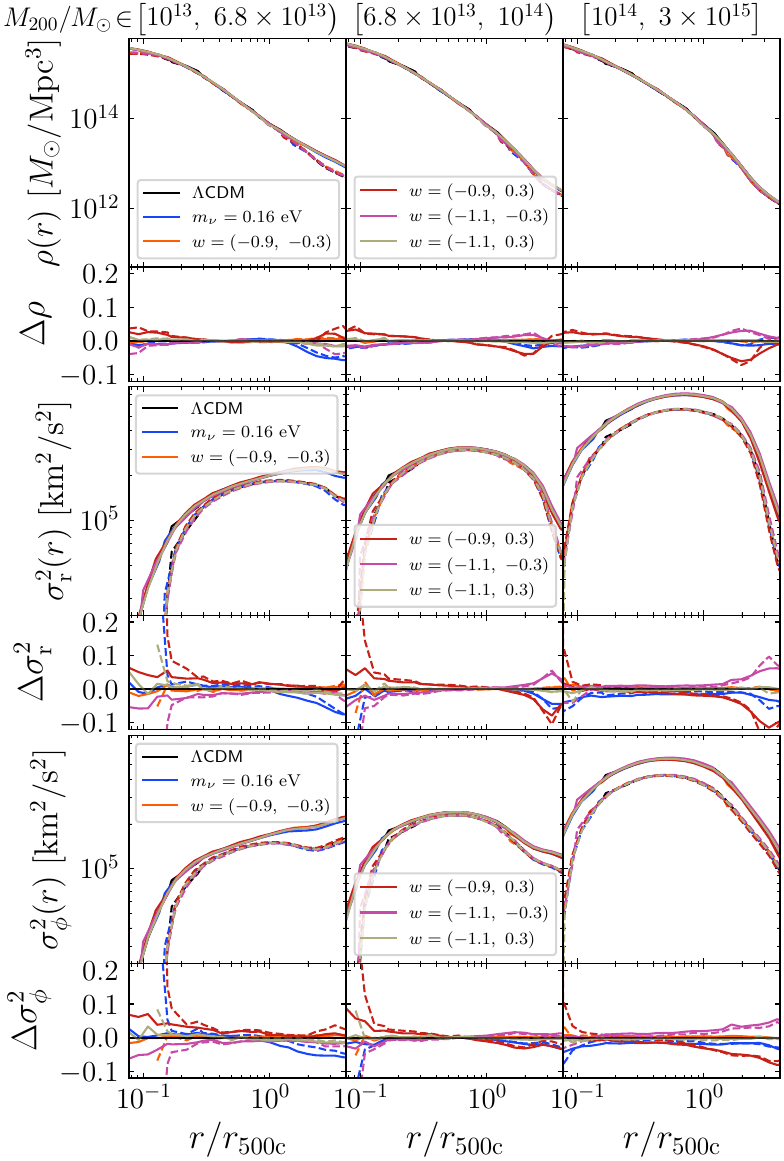}}
    \caption{Stacked \DEMNUni 3D profiles, including the velocity dispersion spherical components $\sigma_{\rm r, \ \theta, \ \phi}$ and the velocity anisotropy $\beta$. Line-style is the same as in Fig.~\ref{fig:profiles_2Dx}. Deviations from \lcdm are more apparent in the mass and velocity dispersion profiles. The most significant effects correspond to the $(w_0, w_a) = (-0.9, 0.3)$ and $(w_0, w_a) = (-1.1, -0.3)$. The cosmic acceleration induced by the former is smaller than \lcdm, thus boosting structure formation and the mass profile. Instead, the latter suppresses the halo growth, leading to smaller profiles. Neutrinos also suppress structure formation, which is most notable in the high-mass bin profiles. }
    \label{fig:DEMNUni_stacked_profiles_3D}
\end{figure*}

\begin{figure*}[h]
    \centering
    \centerline{
    \includegraphics[width=0.41\hsize]{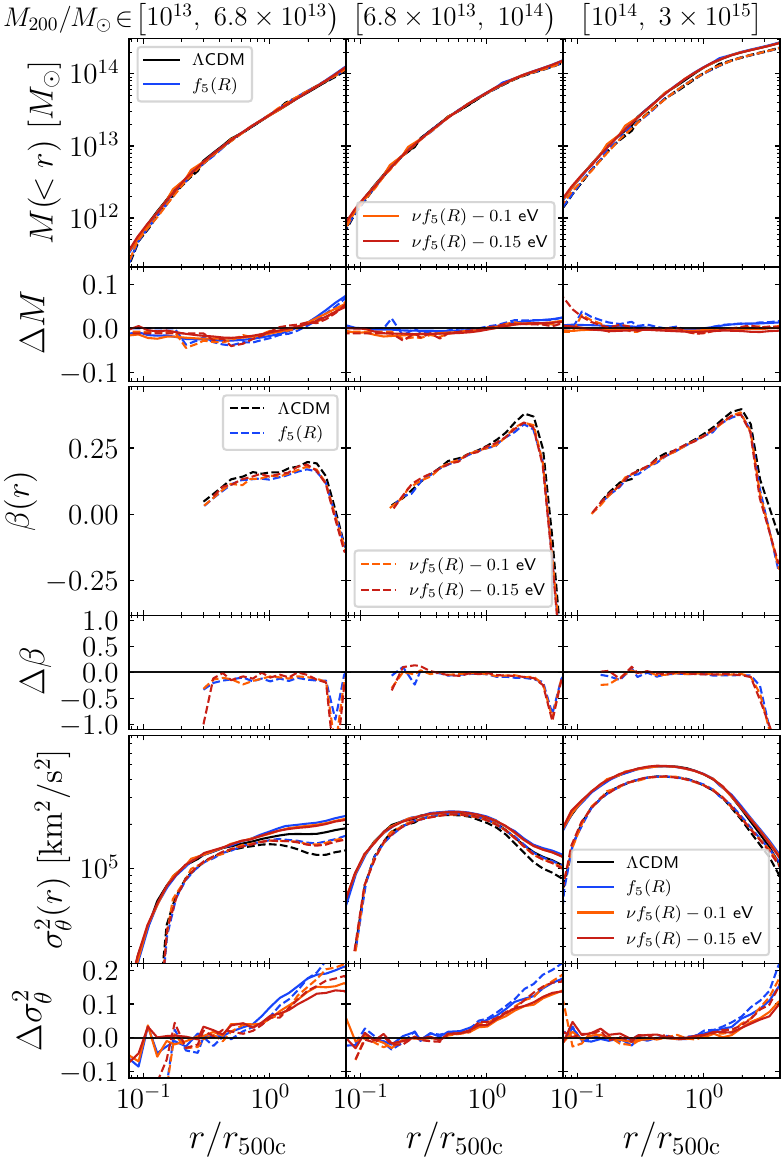}
    \includegraphics[width=0.41\hsize]{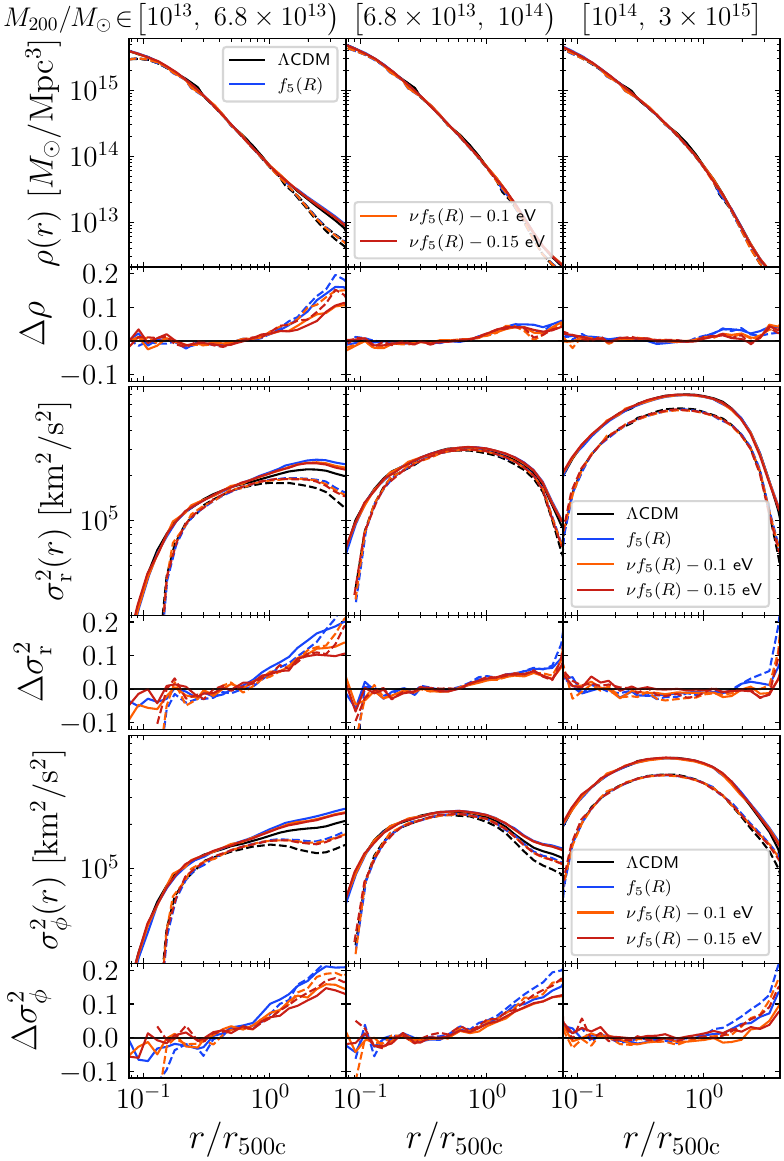}}
    \caption{Stacked \DUSTGRAIN 3D profiles. Line-style is the same as in Fig.~\ref{fig:profiles_2Dx}. The base $f_5(\mathcal{R})$ model results in more significant effects at large radii with respect to the neutrino cases, because neutrinos provide additional screening at their clustering distance. The effect of $f_R$ is also much more significant at large radii in the velocity dispersions when compared to the dynamical dark energy cosmologies of \DEMNUni.}
    \label{fig:DUSTGRAIN_stacked_profiles_3D}
\end{figure*}

In Figs.~\ref{fig:DEMNUni_stacked_profiles_3D} and \ref{fig:DUSTGRAIN_stacked_profiles_3D} we report the three-dimensional mass, density, velocity dispersions and velocity anisotropy profiles for the \DEMNUni and \DUSTGRAINPATHFINDER suites, respectively. 

\newpage

\section{Number of stacked clusters required for significant detection of non-standard effects}
\label{app:detectability}
In Table~\ref{tab:N_req_table} we show the required number of stacked clusters in order to distinguish our non-standard cosmologies achieving a signal-to-noise ratio equal to at least three ($3\sigma$ detection). 

\begin{table*}[h]
 \centering
\caption{\label{tab:N_req_table} Number of stacked cluster required to distinguish a given \DUSTGRAINPATHFINDER (left) or \DEMNUni (right) cosmology at a given redshift with a signal-to-noise ratio (~\ref{eq:SN_ratio}) equal to three. We consider $f_{\rm int} = 0.2$ and $f_{\rm sys} = 0.03$, and only report the $f_{\rm sys} = 0.05$ case if $N_{\rm eq}$ is finite. Moreover, we only compute $N_{\rm req}$ for the median profiles in our lowest mass interval, as they exhibit the strongest non-standard deviations compared to the other mass bins.}
 \begin{tabular}{cccc}
 \hline
 \noalign{\vskip 3pt}
 {Model} & ${z}$ & ${f_{\rm sys}}$ & ${N_{\rm req}}$\\
 \noalign{\vskip 3pt} 
 \hline
 \noalign{\vskip 3pt}
 $f_5(\mathcal{R})$ & 1.1 & 0.03 & $1.3 \times 10^5$ \\ 
 \noalign{\vskip 3pt}
 \hline
 \noalign{\vskip 3pt}
 $\nu f_5(\mathcal{R}) - 0.1\ {\rm eV}$ & 1.1 & 0.03 & $8.5 \times 10^5$ \\ 
 \noalign{\vskip 3pt}
 \hline
 \noalign{\vskip 3pt}
 $\nu f_5(\mathcal{R}) - 0.15\ {\rm eV}$ & 1.1 & 0.03 & Sys. limited\\ 
 \noalign{\vskip 3pt}
 \hline
 \noalign{\vskip 3pt}
 $f_5(\mathcal{R})$ & 0.5 & 0.03 & $9.8 \times 10^3$ \\ 
 \noalign{\vskip 3pt}
 \hline
 \noalign{\vskip 3pt}
 $\nu f_5(\mathcal{R}) - 0.1\ {\rm eV}$ & 0.5 & 0.03 & $5.1 \times 10^4$ \\ 
 \noalign{\vskip 3pt}
 \hline
 \noalign{\vskip 3pt}
 $\nu f_5(\mathcal{R}) - 0.15\ {\rm eV}$ & 0.5 & 0.03 & Sys. limited \\
 \noalign{\vskip 3pt}
 \hline
 \noalign{\vskip 3pt}
 $\nu f_5(\mathcal{R})$ & 0.3 & 0.03 & $1.7 \times 10^3$ \\
 \noalign{\vskip 3pt}
 \hline
 \noalign{\vskip 3pt}
 $\nu f_5(\mathcal{R})$ & 0.3 & 0.05 & $2.9 \times 10^3$ \\ 
 \noalign{\vskip 3pt}
 \hline
 \noalign{\vskip 3pt}
 $\nu f_5(\mathcal{R}) - 0.1\ {\rm eV}$ & 0.3 & 0.03 & $2.2 \times 10^3$ \\
 \noalign{\vskip 3pt}
 \hline
 \noalign{\vskip 3pt}
 $\nu f_5(\mathcal{R}) - 0.1\ {\rm eV}$ & 0.3 & 0.05 & $4.3 \times 10^3$ \\ 
 \noalign{\vskip 3pt}
 \hline
 \noalign{\vskip 3pt}
 $\nu f_5(\mathcal{R}) - 0.15\ {\rm eV}$ & 0.3 & 0.03 & $2.9 \times 10^3$ \\ 
 \noalign{\vskip 3pt}
 \hline
 \noalign{\vskip 3pt}
 $\nu f_5(\mathcal{R}) - 0.15\ {\rm eV}$ & 0.3 & 0.05 & $8.2 \times 10^3$ \\ 
 \noalign{\vskip 3pt}
 \hline
 \end{tabular}
 \hspace{30pt}
 \begin{tabular}{cccc}
 \hline
 \noalign{\vskip 3pt}
 {Model} & ${z}$ & ${f_{\rm sys}}$ & ${N_{\rm req}}$\\
 \noalign{\vskip 3pt} 
 \hline
 \noalign{\vskip 3pt}
 $m_\nu = 0.16\ {\rm eV}$ & 1.1 & 0.03 & Sys. limited \\ 
 \noalign{\vskip 3pt}
 \hline
 \noalign{\vskip 3pt}
 $w = (-0.9,\ -0.3)$ & 1.1 & 0.03 & Sys. limited \\ 
 \noalign{\vskip 3pt}
 \hline
 \noalign{\vskip 3pt}
 $w = (-0.9,\ +0.3)$ & 1.1 & 0.03 & $1.0 \times 10^6$\\ 
 \noalign{\vskip 3pt}
 \hline
 \noalign{\vskip 3pt}
 $w = (-1.1,\ -0.3)$ & 0.5 & 0.03 & Sys. limited \\ 
 \noalign{\vskip 3pt}
 \hline
 \noalign{\vskip 3pt}
 $w = (-1.1,\ +0.3)$ & 0.5 & 0.03 & Sys. limited \\
 \noalign{\vskip 3pt}
 \hline
 \end{tabular}
\end{table*}

\section{Concentration--mass relation from the \DEMNUni and \DUSTGRAINPATHFINDER mass fits} \label{app:cm_appendix}
\begin{figure*}[h]
    \centering
    \includegraphics[width=\linewidth]{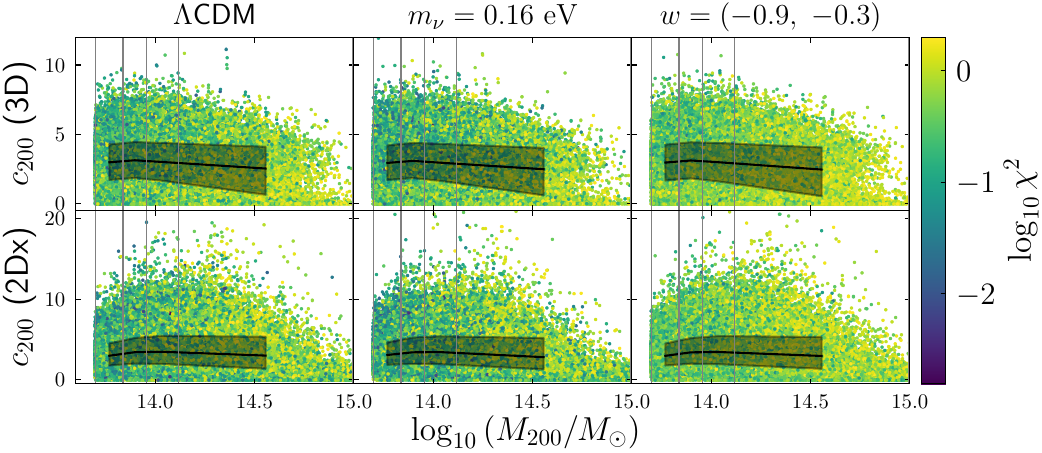}
    \includegraphics[width=\linewidth]{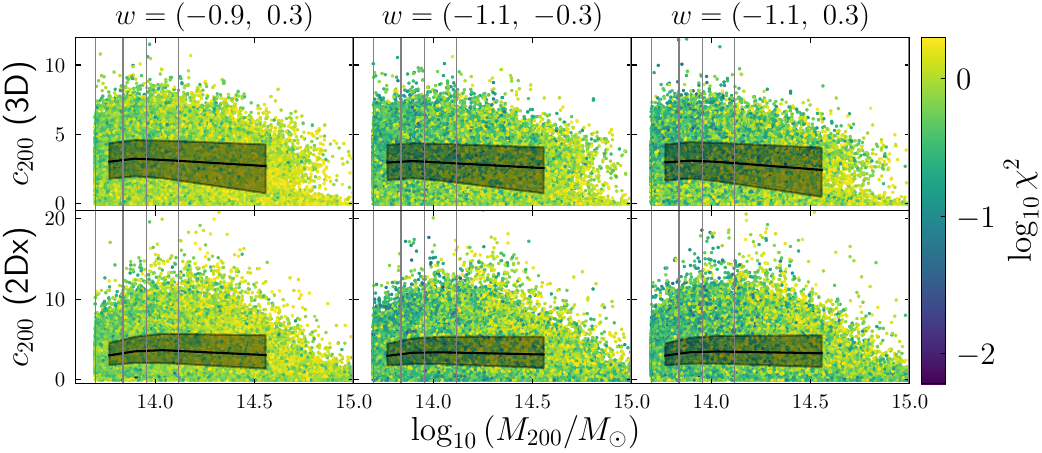}
    \caption{3D and 2Dx concentration--mass relation from the NFW mass fits in each \DEMNUni cosmology analyzed, showing the median value from of the concentration in each mass bin (the black line) and the corresponding $68\,\%$ scatter (the shaded region). The solid grey vertical lines correspond to the edges of the each mass bin.}
    \label{fig:DEMNUni_cM_relation}
\end{figure*}

\begin{figure*}[h]
    \centering
    \includegraphics[width=0.85\linewidth]{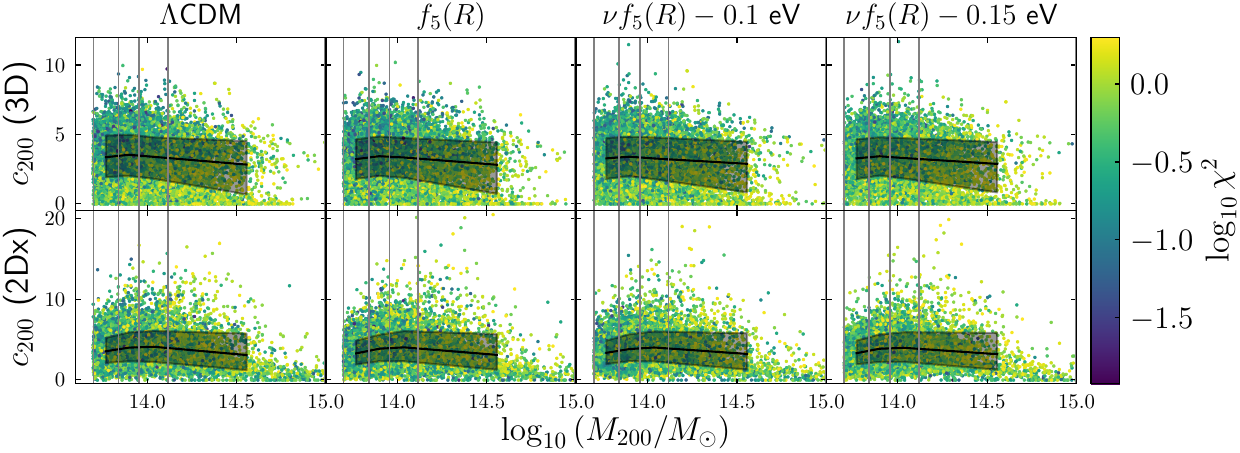}
    \caption{3D and 2Dx concentration--mass relation from the NFW mass fits in each \DUSTGRAINPATHFINDER cosmology analyzed, showing the median value from of the concentration in each mass bin (the black line) and the corresponding $68\,\%$ scatter (the shaded region). The solid grey vertical lines correspond to the edges of the each mass bin.}
    \label{fig:DUSTGRAIN_cM_relation}
\end{figure*}

In Figs.~\ref{fig:DEMNUni_cM_relation} and \ref{fig:DUSTGRAIN_cM_relation} we show the concentration--mass relation for every cosmology considered in \DEMNUni and \DUSTGRAINPATHFINDER respectively. The virial masses and concentrations were obtained from the NFW mass fits we performed. We divided the halo samples in equally-spaced logarithmic mass bins, and the median values of the concentrations in each bin along with their $68\,\%$ scatter are shown in each plot. The grey vertical lines are the mass bin edges.

\section{Velocity anisotropy profiles and concentration--mass relation from \DUSTGRAINPATHFINDER at \texorpdfstring{$z = 0.5$}{z05} and \texorpdfstring{$z = 0.3$}{z03}} \label{app:low_z}
\begin{figure*}[h]
    \centering
    \centerline{
    \includegraphics[width=0.46\linewidth]{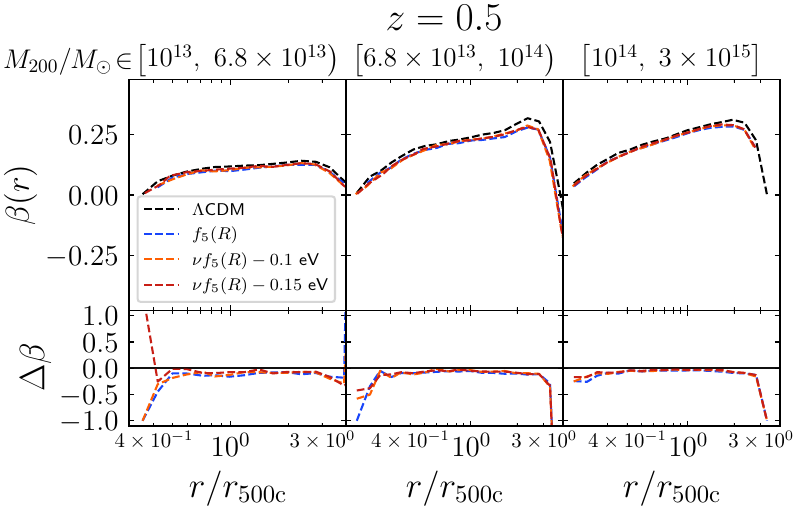}
    \includegraphics[width=0.46\linewidth]{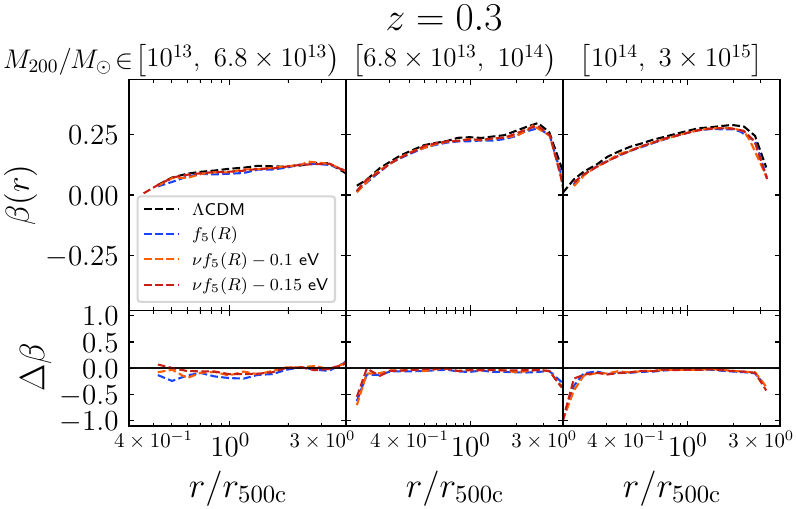}}
    \caption{Velocity anisotropy profiles for two snapshots of the \DUSTGRAINPATHFINDER suite, at $z = 0.5$ (left) and $z = 0.3$ (right). Each colour corresponds to a different model, and the relative differences are computed between that same model and \lcdm. The black dot-dashed lines delimit the $68\,\%$ scatter of the \lcdm model. We also define $\Delta \beta \equiv (\beta - \beta_\mathrm{\Lambda CDM})/\beta_\mathrm{\Lambda CDM}$.}
    \label{fig:DUSTGRAIN_beta_low_z}
\end{figure*}

\begin{figure*}[h]
    \centering
    \centerline{
    \includegraphics[width=0.46\linewidth]{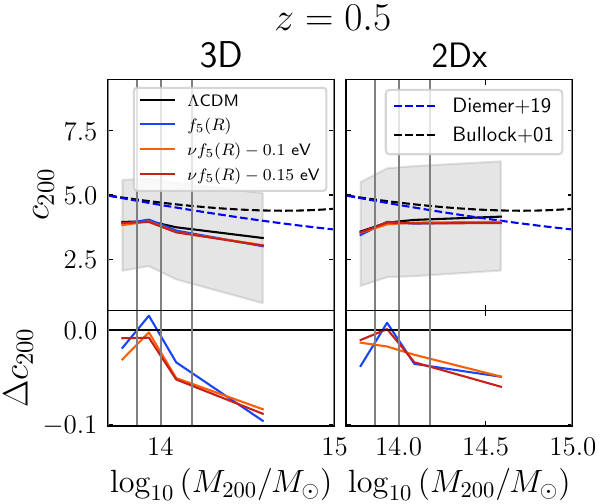}
    \includegraphics[width=0.445\linewidth]{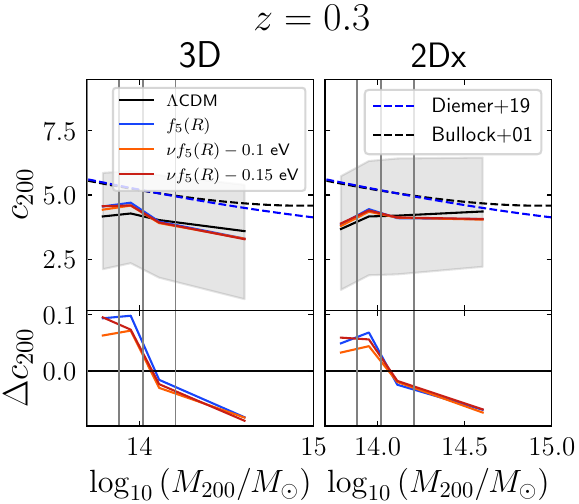}}
    \caption{3D and 2Dx median concentration--mass relations from the NFW mass fits, computed for two snapshots of the \DUSTGRAINPATHFINDER suite, at $z = 0.5$ on the left left, and $z = 0.3$ on the right. The \lcdm scatter is also shown as the shaded region, and the grey vertical lines represent the mass bin edges. We also show the theoretical predictions by~\cite{Bullock_2001} and~\cite{Diemer_2019}, as the black and blue dashed curves, respectively. We define $\Delta c_{200} \equiv (c_{200} - c_\mathrm{200, \Lambda CDM})/c_\mathrm{200, \Lambda CDM}$.}
    \label{fig:DUSTGRAIN_cM_low_z}
\end{figure*}

In Figs.~\ref{fig:DUSTGRAIN_beta_low_z} and \ref{fig:DUSTGRAIN_cM_low_z} we report the velocity anisotropy stacked profiles and the median concentration--mass relation, respectively, computed from the \DUSTGRAINPATHFINDER snapshots at $z = 0.5$ and $z = 0.3$.

\end{appendix}

%\label{LastPage}
\end{document}